\documentclass[11pt]{article}

\usepackage[utf8]{inputenc}
\usepackage[T1]{fontenc}
\usepackage[english]{babel}
\usepackage{geometry}
\usepackage[svgnames,x11names,table]{xcolor} 
\usepackage{graphicx}
\usepackage{amsmath}
\usepackage{amssymb}
\usepackage{url}
\usepackage[colorlinks=true, urlcolor=DarkBlue, linkcolor=black, citecolor=black]{hyperref}
\usepackage{booktabs} 
\usepackage{caption}  
\usepackage{parskip}  
\usepackage{abstract}
\usepackage{fancyhdr} 
\usepackage{fontawesome5} 
\usepackage{enumitem} 
\usepackage{mdframed} 
\usepackage[round, authoryear, comma, sort&compress]{natbib} 
\usepackage{todonotes}

\usepackage{nicematrix} 
\usepackage{tikz} 

\usepackage{array}
\usepackage{multirow}
\usepackage{graphicx}
\usepackage{booktabs}
\usepackage{amssymb}
\usepackage{adjustbox} 
\usepackage{wasysym} 
\usepackage{tabularx}

\usepackage[lining,semibold]{ebgaramond} 
\usepackage[T1]{fontenc} 
\usepackage{libertine} 
\usepackage[varqu,varl]{inconsolata} 
\usepackage{amsmath} 
\usepackage{sectsty}
\allsectionsfont{\sffamily\bfseries} 

\definecolor{MyGreen}{HTML}{C9FFC9}
\definecolor{MyYellow}{HTML}{FFFFC9}
\definecolor{MyRed}{HTML}{FFC9C9}
\definecolor{MyMedium}{HTML}{FFE5C9}

\usepackage{wasysym}
\usepackage{pifont}

\usepackage{enumitem}
\setlistdepth{6}

\newcommand{\impq}{$\bigstar$}
\newcommand{\refbullet}{\ding{43}}

\newcommand{\linkbullet}{$\hookrightarrow$}

\newcommand{\F}{\textsc{F}}
\newcommand{\DEV}{\textsc{DEV}}
\newcommand{\GOV}{\textsc{GOV}}
\newcommand{\NC}{\textsc{NC}}

\setlist[itemize,1]{label=$\bullet$}
\setlist[itemize,2]{label=$\circ$}
\setlist[itemize,3]{label=$\scriptstyle\blacksquare$}
\setlist[itemize,4]{label=$\bullet$}
\setlist[itemize,5]{label=$\bullet$}
\setlist[itemize,6]{label=$\bullet$}
\renewlist{itemize}{itemize}{6}

\fancypagestyle{firstpage}{
    \fancyhf{} 
    \fancyhead[C]{\footnotesize\textit{AI Governance to Avoid Extinction}} 
    \fancyfoot[L]{Contact: \href{mailto:peter@intelligence.org}{peter@intelligence.org}, \  \href{mailto:aaron.scher@intelligence.org}{aaron.scher@intelligence.org}} 
}

\title{AI Governance to Avoid Extinction:\\The Strategic Landscape and Actionable Research Questions}

\author{
    Peter Barnett \and
    Aaron Scher
}

\date{\vspace{-1ex}\textit{Machine Intelligence Research Institute, Technical Governance Team}\\ \vspace{2ex}
May 2025} 

\hypersetup{
    pdftitle={AI Governance to Avoid Extinction: The Strategic Landscape and Actionable Research Questions}, 
    pdfauthor={Peter Barnett, Aaron Scher},
    pdfsubject={AI Governance Research Agenda},
    pdfkeywords={AI Safety, AI Governance, MIRI, Research Agenda},
    breaklinks=true 
}

\newlength{\ColScenario}
\newlength{\ColProsCons}
\newlength{\ColRisk}

\begin{document}

\maketitle
\thispagestyle{firstpage}

\begin{abstract}
Humanity appears to be on course to soon develop AI systems that substantially outperform human experts in all cognitive domains and activities. We believe the default trajectory has a high likelihood of catastrophe, including human extinction. Risks come from failure to control powerful AI systems, misuse of AI by malicious rogue actors, war between great powers, and authoritarian lock-in. This research agenda has two aims: to describe the strategic landscape of AI development and to catalog important governance research questions. These questions, if answered, would provide important insight on how to successfully reduce catastrophic risks.

We describe four high-level scenarios for the geopolitical response to advanced AI development, cataloging the research questions most relevant to each. Our favored scenario involves building the technical, legal, and institutional infrastructure required to internationally restrict dangerous AI development and deployment (which we refer to as an Off Switch), which leads into an internationally coordinated Halt on frontier AI activities at some point in the future. The second scenario we describe is a US National Project for AI, in which the US Government races to develop advanced AI systems and establish unilateral control over global AI development. We also describe two additional scenarios: a Light-Touch world similar to that of today and a Threat of Sabotage situation where countries use sabotage and deterrence to slow AI development.

In our view, apart from the Off Switch and Halt scenario, all of these trajectories appear to carry an unacceptable risk of catastrophic harm. Urgent action is needed from the US National Security community and AI governance ecosystem to answer key research questions, build the capability to halt dangerous AI activities, and prepare for international AI agreements.
\end{abstract}

\pagebreak
\section*{Executive Summary}
\addcontentsline{toc}{section}{Executive Summary}

The default trajectory of AI development has an unacceptably high likelihood of leading to human extinction. Critical research and policy is needed to course correct. This report describes four high-level scenarios for the course of advanced AI development and the geopolitical response to it. The report then catalogs important governance research questions within each of these scenarios. These questions, if answered, would provide important insight on how to successfully reduce catastrophic risks. We believe the main focus of this research should be ensuring humanity's collective ability to restrict dangerous AI development and deployment, up to and including a halt on frontier AI activities, if and when there is sufficient political will to do so.

Humanity appears on course to develop AI systems that exceed human performance at most cognitive tasks very soon. AI companies such as OpenAI, Anthropic, and Google DeepMind are explicitly aiming at this goal, and many experts think they will succeed in the next few years~\citep{todd2025agi, kokotajlo2025ai2027}.  This development alone will be one of the most transformative events in human history---similar in scale to the Industrial Revolution---and the world does not appear prepared. But the most worrisome challenges arise as these systems grow to substantially surpass humanity in all strategically relevant activities, becoming what is often referred to as artificial superintelligence (ASI).

In our view, the field of AI is on track to produce ASI while having little to no understanding of how these systems function and no robust means to steer and control their behavior. The coming years and decades thus present major challenges if we are to avoid large-scale risks from advanced AI systems, including:
\begin{itemize}
    \item \textbf{Loss of Control}: AI disempowers humanity in pursuit of goals not aligned with our collective interests, which likely results in human extinction. \href{https://www.safe.ai/work/statement-on-ai-risk}{Many experts}~\citep{cais_statement_2023} have expressed concerns of this type. We see this as the \textit{default} outcome of near-term ASI and refer readers to previous work on the topic: \href{https://arxiv.org/abs/2310.17688}{Managing extreme AI risks amid rapid progress}~\citep{bengio2024managing}, \href{https://www.alignmentforum.org/posts/pRkFkzwKZ2zfa3R6H/without-specific-countermeasures-the-easiest-path-to}{Without specific countermeasures}~\citep{cotra_without_2022}, \href{https://arxiv.org/abs/2209.00626}{The Alignment Problem from a Deep Learning Perspective}~\citep{ngo_alignment_2025}, \href{https://intelligence.org/2015/07/24/four-background-claims/}{Four Background Claims}~\citep{soares2015four}, \href{https://intelligence.org/wp-content/uploads/2024/12/Misalignment_and_Catastrophe.pdf}{Without fundamental advances}~\citep{barnett_without_2024}. 
    \item \textbf{Misuse}: Malicious or incautious actors use AI, deliberately or accidentally, leading to catastrophic harm. This may include biological weapons or other weapons of mass destruction.
    \item \textbf{War}: AI-related conflict between great powers causes catastrophic harm.
    \item \textbf{Authoritarianism/Lock-in}: The world locks in values or conditions that are harmful, such as a stable, global authoritarian regime.
\end{itemize}

This report provides an extensive collection of research questions intended to assist the US National Security community and the AI governance research ecosystem in preventing these catastrophic outcomes.

The presented research agenda is organized by four high-level scenarios for the trajectory of advanced AI development in the coming years: a coordinated global Halt on dangerous AI development (Off Switch and Halt), a US National Project leading to US global dominance (US National Project), continued private-sector development with limited government intervention (Light-Touch), and a world where nations maintain AI capabilities at safe levels through mutual threats of interference (Threat of Sabotage). We explore questions about the stability and viability of the governance strategy that underlies each scenario: What preconditions would make each scenario viable? What are the main difficulties of the scenario? How can these difficulties be reduced, including by transitioning to other strategies? What is a successful end-state for the scenario?
\begin{figure}[thbp]
    \centering
    \includegraphics[width=\textwidth]{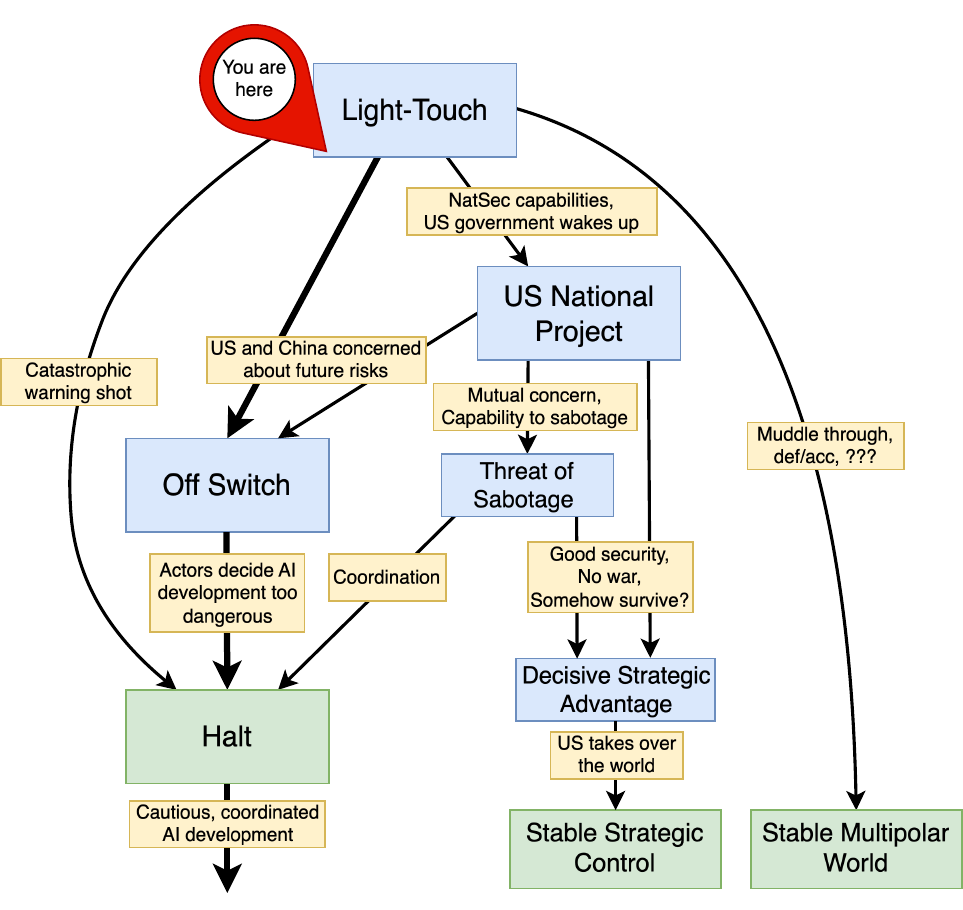} 
    \caption{A rough tree of the scenarios discussed in this report and how AI governance paradigms may evolve. This diagram assumes catastrophe is avoided at each step, so the myriad failures are omitted. This is simplified, and one could draw many more connections between the scenarios.}
    \label{fig:paradigm_tree_exec_summary}
\end{figure}
\subsection*{Off Switch and Halt} \label{sec:exec_summary_off_switch}

The first scenario we describe involves the world coordinating to develop the ability to monitor and restrict dangerous AI activities. Eventually, this may lead to a \textit{Halt}: a global moratorium on the development and deployment of frontier AI systems until the field achieves justified confidence that progress can resume without catastrophic risk. Achieving the required level of understanding and assurance could be extraordinarily difficult, potentially requiring decades of dedicated research.

While consensus may currently be lacking about the need for such a Halt, it should be uncontroversial that humanity should be able to stop AI development in a coordinated fashion if or when it decides to do so. We refer to the technical, legal, and institutional infrastructure needed to Halt on demand as an \textit{Off Switch} for AI.

We are focused on an Off Switch because we believe an eventual Halt is the best way to reduce Loss of Control risk from misaligned advanced AI systems. Therefore, we would like humanity to build the capacity for a Halt in advance. Even for those skeptical about alignment concerns, there are many reasons Off Switch capabilities would be valuable. These Off Switch capabilities---the ability to monitor, evaluate, and, if necessary, enforce restrictions on frontier AI development---would also address a broader set of national security concerns. They would assist with reducing risks from terrorism, geopolitical destabilization, and other societal disruption.

\textbf{This research agenda primarily focuses on the Off Switch and Halt scenario, as we believe a Halt is the most credible path to avoiding human extinction.}

The research questions in this scenario primarily concern the design of an Off Switch, and the key details to enforcing a Halt. For example:
\begin{itemize}[itemsep=-0.1em]
    \item How do we create common understanding about AI risks and get buy-in from different actors to build the Off Switch?
    \item What features are needed for an effective Off Switch, and how can the world implement them?
    \item What are the trends in compute requirements for frontier AI systems?
    \item How can governments use controls on specialized chips to institute a long-term moratorium on dangerous AI?
    \item What is a suitable emergency response plan, for both AI projects and governments? How should actors respond to an AI emergency, including both mitigating immediate harm and learning the right lessons?
    \item What compute needs to be monitored after a Halt is initiated?
    \item How can governments monitor compute they know about, especially to ensure it isn’t being used to violate a Halt?
    \item Other than compute and security, what levers exist to control AI development and deployment?
\end{itemize}

\subsection*{US National Project} \label{sec:exec_summary_usnp}

The second scenario is the US National Project, in which the US government races to develop advanced AI systems and establish unilateral control over global AI development. This scenario is based on stories discussed previously by \href{https://situational-awareness.ai/}{Leopold Aschenbrenner} and Anthropic CEO \href{https://darioamodei.com/machines-of-loving-grace}{Dario Amodei}. A successful US National Project requires navigating numerous difficult challenges:
\begin{itemize}[itemsep=-0.1em]
    \item Maintaining a lead over other countries
    \item Avoiding the proliferation of advanced AI capabilities to terrorists
    \item Avoiding war with other countries
    \item Developing advanced AI capabilities despite potential hardware or software limits
    \item Avoiding the development of misaligned AI systems that lead to Loss of Control.
    \item Converting its AI capabilities advantage into a decisive advantage over other actors.
    \item Avoiding governance failure such as authoritarian power grabs.
\end{itemize}

Some of these challenges look very difficult, such that pursuing the project would be unacceptably dangerous. We encourage other approaches, namely coordinating a global Off Switch and halting dangerous AI development.

The research questions in this section examine how to prepare for and execute on the project, along with approaches for pivoting away from a National Project and into safer strategies. For example:
\begin{itemize}[itemsep=-0.1em]
    \item How can the US lead in AI capabilities be measured?
    \item How could a centralized US National Project bring in other AI development projects (domestic and international)?
    \item What ready-to-go research should the US National Project prioritize using AIs for, when AIs are capable of automating AI safety research?
    \item What is a safety plan that would allow an AI project to either successfully build aligned advanced AI, or safely notice that its development strategy is too dangerous?
    \item What mechanisms are available to reduce racing between nations?
    \item How might the US National Project achieve a decisive strategic advantage using advanced AI?
    \item How could the US National Project recognize that its strategy is too dangerous?
\end{itemize}

\subsection*{Light-Touch} \label{sec:exec_summary_light_touch}

Light-Touch is similar to the current world, where the government takes a light-touch approach to regulating AI companies. We have not seen a credible story for how this situation is stable in the long run. In particular, we expect governments to become more involved in AI development as AIs become strategically important, both militarily and economically. Additionally, the default trajectory will likely involve the open release of highly capable AI models. Such models would drastically increase large-scale risks from malicious actors, for instance, by assisting with biological weapon development. One approach \href{https://vitalik.eth.limo/general/2023/11/27/techno_optimism.html}{discussed previously}~\citep{buterin_my_2023} to remedy this situation is \textit{defensive acceleration}: investing heavily in defense-oriented technologies in order to combat offensive use. We are pessimistic about such an approach because some emerging technologies---such as biological weapons---appear much easier to weaponize than to defend against. The Light-Touch approach also involves similar risks to those in the US National Project, such as misalignment and war. We think the Light-Touch scenario is extremely unsafe.

The research questions in this section are largely about light government interventions to improve the situation or transitioning into an Off Switch strategy or US National Project. For example:
\begin{itemize}[itemsep=-0.1em]
    \item What light-touch interventions are available to coordinate domestic AI projects to reduce corporate race dynamics?
    \item What dangerous capabilities will the government care about solely controlling, and when might these be developed?
    \item How does development of national security technology by the private sector typically work? What are the most important lessons to take away from existing public-private partnerships for such technologies?
    \item What kinds of transparency should governments have into private AI development?
    \item How can AI developers implement strong security for AI model weights? How could the government promote this?
    \item How can AI developers implement strong security for algorithmic secrets? How could the government promote this?
\end{itemize}

\subsection*{Threat of Sabotage} \label{sec:exec_summary_sabotage}

AI progress could disrupt the balance of power between nations (e.g., enable a decisive military advantage), so countries might take substantial actions to interfere with advanced AI development. Threat of Sabotage, similar to Mutual Assured AI Malfunction (MAIM) described in \textit{\href{https://www.nationalsecurity.ai/}{Superintelligence Strategy}} ~\citep{hendrycks_superintelligence_2025}, describes a strategic situation where AI development is slow because countries threaten to sabotage rivals’ AI progress. Actual sabotage may occur, via substantial actions to interfere with AI development, although merely the threat of sabotage could be sufficient to keep AI progress slow. The state of thinking about this scenario is nascent, and we are excited to see further analysis of its viability and implications.

One of our main concerns is that the situation only remains stable if there is a high degree of visibility into AI projects and potential for sabotage, but these are both complicated factors that are difficult to predict in advance. Visibility and potential for sabotage are both likely high in the current AI development regime, where frontier AI training requires many thousands of advanced chips, but this situation could change.

The research questions in this section focus on better understanding the viability of the scenario and transitioning into a more cooperative Off Switch scenario. For example:
\begin{itemize}[itemsep=-0.1em]
    \item Will the security levels in AGI projects be at the required level to enable the Threat of Sabotage dynamic? Threat of Sabotage largely requires that security in the main AGI projects is strong enough to prevent proliferation to non-state actors, but weak enough to enable countries to see each other’s progress and sabotage each other.
    \item What are the key methods countries might use to sabotage each other’s AGI projects? How effective are these? Would these prompt further escalation?
    \item How long would it take an actor to accomplish various key AI activities given different starting capabilities? For example, how long would it take to domestically produce AI chips, build an AI data center, or reach a particular AI model capability level?
    \item What might enable a transition from a Threat of Sabotage regime to an international Off Switch-style agreement with verification? For example, mechanisms for credible non-aggression or benefit sharing.
\end{itemize}

\subsection*{Understanding the World} \label{sec:exec_summary_understanding}

There are some research projects that are generally useful for understanding the strategic situation and gaining situational awareness. We include some research projects in this section because they are broadly useful across many AI development trajectories. For example:
\begin{itemize}[itemsep=-0.1em]
    \item How viable is compute governance?
    \item Understanding and forecasting model capabilities
    \item What are the trends in the cost of AI inference?
    \item What are the implications of the inference scaling regime?
    \item What is the state of AI hardware and the computing stock?
    \item What high-level plans and strategies for AI governance seem promising?
\end{itemize}

\subsection*{Outlook} \label{sec:exec_summary_conclusion}

Humanity is on track to soon develop AI systems smarter than the smartest humans; this might happen in the 2020s or 2030s. On both the current trajectory and some of the most likely variations (such as a US National Project), there is an unacceptably large risk of catastrophic harm. Risks include terrorism, world war, and risks from AI systems themselves (i.e., loss of control due to AI misalignment).

Humanity should take a different path. We should build the technical, legal, and institutional infrastructure needed to halt AI development if and when there is political will to do so. This Halt would provide the conditions for a more mature AI field to eventually develop this technology in a cautious, safe, and coordinated manner.

\begin{table}[htbp]
\centering
\caption{The main pros and cons of the scenarios and how much of each core risk they involve. Ratings are based on our analysis of the scenarios. Color coding reflects the risk level. (Same as Table \ref{tab:scenario_ratings_intro})}
\label{tab:scenario_ratings_exec_summary}
\begin{NiceTabular}{@{} >{\raggedright\arraybackslash}p{\ColScenario} 
                   >{\arraybackslash}p{\ColProsCons} 
                   >{\arraybackslash}p{\ColProsCons} 
                   >{\centering\arraybackslash}m{\ColRisk}      
                   >{\centering\arraybackslash}m{\ColRisk}      
                   >{\centering\arraybackslash}m{\ColRisk}      
                   >{\centering\arraybackslash}m{\ColRisk}      
                }[cell-space-limits = 4pt] 
\toprule
Scenario & Pros & Cons & \Block[c]{}{\mbox{Loss of}\\Control} & Misuse & War & \Block[c]{}{Bad\\\mbox{Lock-in}} \\
\hline

Off Switch and Halt &
    \parbox[t]{\hsize}{%
    \raggedright 
    \begin{itemize}[leftmargin=*, label=\textbullet, nosep, topsep=0pt, itemsep=0pt, partopsep=0pt, parsep=0pt] 
        \item Careful AI \mbox{development}
        \item International \mbox{legitimacy}
        \item Slow societal \mbox{disruption}
    \end{itemize}
    } & 
    \parbox[t]{\hsize}{%
    \raggedright 
    \begin{itemize}[leftmargin=*, label=\textbullet, nosep, topsep=0pt, itemsep=0pt, partopsep=0pt, parsep=0pt]
        \item Difficult to implement
        \item Not a complete \mbox{strategy}
        \item Slow AI benefits
    \end{itemize}%
    } &
    \Block[fill=MyGreen]{1-1}{Low} & \Block[fill=MyGreen]{1-1}{Low} & \Block[fill=MyGreen]{1-1}{Low} & \Block[fill=MyMedium]{1-1}{Mid} \\
\hline
US National Project &
    \parbox[t]{\hsize}{%
    \raggedright 
    \begin{itemize}[leftmargin=*, label=\textbullet, nosep, topsep=0pt, itemsep=0pt, partopsep=0pt, parsep=0pt]
        \item Centralized ability to implement safeguards
        \item Limited proliferation
    \end{itemize}%
    } &
    \parbox[t]{\hsize}{%
    \raggedright 
    \begin{itemize}[leftmargin=*, label=\textbullet, nosep, topsep=0pt, itemsep=0pt, partopsep=0pt, parsep=0pt]
        \item Arms race
        \item Breaking international norms
    \end{itemize}%
    } &
    \Block[fill=MyRed]{1-1}{High} & \Block[fill=MyGreen]{1-1}{Low} & \Block[fill=MyRed]{1-1}{High} & \Block[fill=MyRed]{1-1}{High} \\
\hline
Light-Touch &
    \parbox[t]{\hsize}{%
    \raggedright 
    \begin{itemize}[leftmargin=*, label=\textbullet, nosep, topsep=0pt, itemsep=0pt, partopsep=0pt, parsep=0pt]
        \item Fast economic benefit
        \item Less international provocation
        \item Easy to implement (default)
    \end{itemize}%
    } &
    \parbox[t]{\hsize}{%
    \raggedright 
    \begin{itemize}[leftmargin=*, label=\textbullet, nosep, topsep=0pt, itemsep=0pt, partopsep=0pt, parsep=0pt]
        \item Corporate racing
        \item Proliferation
        \item Limited controls \mbox{available}
        \item Untenable
    \end{itemize}%
    } &
    \Block[fill=MyRed]{1-1}{High} & \Block[fill=MyRed]{1-1}{High} & \Block[fill=MyMedium]{1-1}{Mid} & \Block[fill=MyMedium]{1-1}{Mid} \\
\hline
Threat of Sabotage &
    \parbox[t]{\hsize}{%
    \raggedright 
    \begin{itemize}[leftmargin=*, label=\textbullet, nosep, topsep=0pt, itemsep=0pt, partopsep=0pt, parsep=0pt]
        \item Slower AI development
        \item Limited cooperation needed
    \end{itemize}%
    } &
    \parbox[t]{\hsize}{%
    \raggedright 
    \begin{itemize}[leftmargin=*, label=\textbullet, nosep, topsep=0pt, itemsep=0pt, partopsep=0pt, parsep=0pt]
        \item Ambiguous stability
        \item Escalation
    \end{itemize}%
    } &
    \Block[fill=MyMedium]{1-1}{Mid} & \Block[fill=MyGreen]{1-1}{Low} & \Block[fill=MyRed]{1-1}{High} & \Block[fill=MyMedium]{1-1}{Mid} \\
\bottomrule 

\CodeAfter
  \tikz \draw [gray] 
    (2-|4) -- (6-|4)  
    (2-|5) -- (6-|5)  
    (2-|6) -- (6-|6)  
    (2-|7) -- (6-|7)  
    (2-|8) -- (6-|8); 
\end{NiceTabular}
\end{table}


\clearpage
\tableofcontents
\clearpage

\section{Introduction} \label{sec:introduction}

\subsection{Who we are} \label{sec:who_we_are}

The Machine Intelligence Research Institute (MIRI) is a research nonprofit based in Berkeley, California, founded in 2000. We focus on increasing the probability that humanity can safely navigate the transition to a world with smarter-than-human AI. Our primary concern is mitigating catastrophic and extinction risks associated with artificial intelligence systems as they approach and surpass human capabilities.

MIRI’s initial work was in AI safety and alignment technical research and field building. Starting in mid-2023 and announced in early \href{https://intelligence.org/2024/01/04/miri-2024-mission-and-strategy-update/}{2024}, we have pivoted our focus more towards policy and communications objectives~\citep{bourgon2024miri}.

We are worried that humanity is on course to develop advanced AI systems without first solving critical safety problems, such as ensuring these systems have goals that are compatible with humanity’s survival---something the field does not know how to do. Failure to solve such safety problems in time could lead to catastrophic harm up to and including human extinction. The world, including the public and policymakers, is extremely unprepared to deal with future rapid AI developments. Humanity should not build AI systems that could lead to extinction before we have justified confidence in a positive outcome.

\subsection{About this document} \label{sec:about_this_doc}

This research agenda lays out a few high-level scenarios for the geopolitical response to advanced AI development in the coming years. These scenarios are partially built on writing by others; however, previous articulations are often vague. In the process of interpreting such scenarios, we may have misrepresented others’ positions. We welcome clarification and encourage more effort on high-level stories with positive outcomes---the world is in need of detailed plans that might actually work.

These scenarios serve as a guide to catalog about 400 research questions that, with better answers, could help humanity navigate the AI transition and avoid catastrophic risk. This list covers both technical and non-technical (e.g., institutional) questions. We have tried to make these questions somewhat concrete, but many are still one or two steps removed from well-scoped research projects. We have cited some relevant prior work, but this should not be considered a comprehensive review of the literature for all listed projects. The previous work most similar to this agenda is \textit{Analysis of Global AI Governance Strategies}~\citep{martin_analysis_2024} and \textit{Open Problems in Technical AI Governance}~\citep{reuel_open_2024}. As appendices, we have included a \hyperref[sec:glossary]{glossary}, and a \hyperref[sec:appendix_disagreements]{discussion of core assumptions}.

We hope this agenda can serve as a guide to others working on reducing large-scale AI risks. While we are likely to pick our future projects from this list of open research questions, there is far too much to be done. We will make progress on only a small subset of these questions, and we encourage other researchers to work on these questions as well---please reach out if coordination could be useful. \textbf{This document reflects the views of the Technical Governance Team, rather than MIRI overall.}

\subsection{Core considerations} \label{sec:core_considerations}

This section discusses the main risks we are interested in avoiding, factors that exacerbate or reduce these risks, and other key background claims about AI development. These considerations affect how we evaluate various high-level plans for advanced AI development.

\subsubsection{Core risks to avoid} \label{sec:core_risks}
\begin{itemize}
    \item \textbf{Loss of Control}: AI disempowers humanity in pursuit of goals not aligned with our collective interests, which likely results in human extinction. \href{https://www.safe.ai/work/statement-on-ai-risk}{Many experts}~\citep{cais_statement_2023} have expressed concerns of this type. We see this as the \textit{default} outcome of near-term artificial superintelligence (ASI), and refer readers to previous work on the topic: \href{https://arxiv.org/abs/2310.17688}{Managing extreme AI risks amid rapid progress}~\citep{bengio2024managing}, \href{https://www.alignmentforum.org/posts/pRkFkzwKZ2zfa3R6H/without-specific-countermeasures-the-easiest-path-to}{Without specific countermeasures}~\citep{cotra_without_2022}, \href{https://arxiv.org/abs/2209.00626}{The Alignment Problem from a Deep Learning Perspective}~\citep{ngo_alignment_2025}, \href{https://intelligence.org/2015/07/24/four-background-claims/}{Four Background Claims}~\citep{soares2015four}, \href{https://intelligence.org/wp-content/uploads/2024/12/Misalignment_and_Catastrophe.pdf}{Without fundamental advances}~\citep{barnett_without_2024}. 
    \item \textbf{Misuse}: Malicious or incautious actors use AI, deliberately or accidentally, leading to catastrophic harm. This may include biological weapons or other weapons of mass destruction.
    \item \textbf{War}: AI-related conflict between great powers causes catastrophic harm.
    \item \textbf{Authoritarianism/Lock-in}: The world locks in values or conditions that are harmful, such as a stable, global authoritarian regime.
\end{itemize}
These are the main criteria we are using to assess the desirability of different AI development strategies---we would like to avoid strategies that impose substantial risk of catastrophic harm.

\subsubsection{Factors that worsen these risks} \label{sec:factors_worsen}
\begin{itemize}
    \item \textbf{Racing}: State actors and companies rapidly develop powerful AI due to perceived geopolitical, security, or commercial interests. Racing reduces one’s ability to exercise caution, perform necessary alignment research, or notice signs of danger.
    \item \textbf{Proliferation}: Powerful, dual-use, AI technology is accessible to a wide range of actors (including terrorists and rogue states). These actors could \textit{misuse} AI to cause catastrophic harm. Even if particular AI systems are not directly useful for harmful ends, their proliferation is still dangerous if they could assist with AI R\&D. Proliferation of such AI systems increases \textit{loss of control} risk because it increases the number of actors who can do frontier AI development, making coordinated pausing of AI development more difficult.
    \item \textbf{Dangerous New Tech}: Advanced AI enables the development of new technologies that the world is not prepared for (e.g., nanotech, weapons that undermine nuclear deterrence).
\end{itemize}

\subsubsection{Factors that reduce these risks} \label{sec:factors_reduce}
\begin{itemize}
    \item \textbf{Global Cooperation}: Nations can cooperate, avoid racing, and make agreements to share in the benefits of AI and prevent bad outcomes.
    \item \textbf{Security}: AI development projects have strong security around model weights and algorithms, making it difficult for state and non-state actors to steal critical intellectual property. Such theft would cause proliferation and thus worsen racing by allowing more actors to compete at the frontier.
    \item \textbf{Ability to Slow Down or Halt}: Humanity retains the ability to slow down or halt AI development---likely requiring substantial coordination---in order to avoid large-scale danger. To be successful, humanity needs the ability to impose restrictions on AI development, and ideally to robustly verify and enforce those restrictions.
    \item \textbf{AI Alignment Research}: AI developers understand the AI systems they build, well enough to shape these systems’ goals.
\end{itemize}

\subsubsection{Other background claims} \label{sec:background_claims}
\begin{itemize}
    \item \textbf{Capabilities}: On the default trajectory, AI systems will become extremely capable, far surpassing human experts across many domains. In addition to being faster and smarter than humans, they will also be capable of accelerating the development of dangerous technologies like biological weapons.
    \item \textbf{Autonomy}: Future AI systems will operate with increasing independence, not merely as tools. Competitive pressures will drive deployment of systems that can pursue long-term goals with minimal human oversight.
    \item \textbf{AI Misalignment}: When systems become both autonomous and highly capable, we face serious risks. The scientific fields of machine learning and AI alignment have an extremely limited understanding of how frontier AI systems work and what goals they develop during training. With the fields’ current understanding, we cannot properly inspect their underlying motivations---and do not even know how to do so in principle. Deploying such systems would pose an unacceptable risk of humanity losing control. We do not thoroughly argue about the likelihood of loss of control from misalignment in this research agenda, and instead refer readers to previous work on the topic: \href{https://arxiv.org/abs/2209.00626}{The Alignment Problem from a Deep Learning Perspective}~\citep{ngo_alignment_2025}, \href{https://intelligence.org/wp-content/uploads/2024/12/Misalignment_and_Catastrophe.pdf}{Without fundamental advances}~\citep{barnett_without_2024}, \href{https://www.alignmentforum.org/posts/uMQ3cqWDPHhjtiesc/agi-ruin-a-list-of-lethalities}{AGI Ruin: A List of Lethalities}~\citep{yudkowsky2022agi}. 
    \item \textbf{Alignment Difficulty:} AI Alignment is a hard technical problem, and it is unclear exactly how hard it is. Making the necessary research progress to robustly shape advanced AI systems’ goals may take decades of dedicated effort.
    \item \textbf{Timelines}: AI systems capable of posing catastrophic risk are likely to be developed in the coming years or decades, potentially in the 2020s.
\end{itemize}
These points shape our strategic thinking: avoid the dire risks, reduce the factors that accelerate them, and build up the mechanisms (cooperation, security, slow-down options) to keep humanity in control of AI’s future.

\subsection{Halting dangerous AI development} \label{sec:halting_dev}

The development of smarter-than-human AI is extremely dangerous, and we along with \href{https://www.safe.ai/work/statement-on-ai-risk}{many experts believe}~\citep{cais_statement_2023} it could cause human extinction. Therefore, we believe humanity should take a different approach: halting frontier AI development before it reaches dangerous levels, and spending as much time as necessary to ensure safe development. This would be a global \textit{Halt} on frontier AI development and deployment: a coordinated and collectively enforced moratorium on dangerous AI, potentially lasting decades. This moratorium would be maintained until there are robust solutions to the technical problems required to ensure powerful AI systems will not cause catastrophe. Given the current state of the field, we are not yet aware of all of these technical problems, let alone able to solve them.

At the present time, there is not sufficient consensus or political will to implement a halt to frontier AI development. Therefore, our main ask is that humanity \textbf{preserve the option to halt}. We conceptualize this optionality as an \textit{Off Switch} for AI. An Off Switch means establishing the technical, legal, and institutional infrastructure required to shut down sufficiently dangerous AI projects on the global level, if and when it is deemed necessary. Building this capacity does not entail shutting down immediately, and it does not require that everybody agree with any particular assessment of the risks.

Building Off Switch capabilities would help address a broader set of national security risks beyond Loss of Control. The solutions needed to monitor, evaluate, and if necessary enforce restrictions on frontier AI development will be critical to address risks from terrorism, proliferation of dangerous capabilities, and unauthorized use. We believe many parties should support building an Off Switch, even those who disagree with us about catastrophic risk from misaligned AI. While a coordinated Off Switch would pose some risk of being abused for authoritarian ends, careful design and multilateral enforcement can mitigate this risk.

\textbf{An Off Switch encompasses both the capacity to shut down AI systems in response to urgent risks and the ability to implement a global moratorium}. In this document, the capacity for rapidly responding to urgent risks (such as rogue AI systems or AI projects) falls under \hyperref[sec:acute_risk]{Acute risk mitigation}. While many stakeholders support acute risk mitigation measures, these alone are insufficient. A substantive halt is likely necessary for the field to mature enough for AI progress to proceed safely, if given appropriate oversight. The capacity for a sustained long-term halt is key to the Off Switch’s success. The sections \hyperref[sec:short_term_pause]{Short-term pause} and \hyperref[sec:long_term_moratorium]{Long-term moratorium} cover capacities which are needed once the most urgent threats have been dealt with.

This document partially focuses on research that is key to realizing a global Halt, especially via building an Off Switch in advance. It also discusses other high-level scenarios for AI development which we think are relatively likely, including how they interact with a global Halt. Unfortunately, these other scenarios often appear very dangerous, e.g., a national AI project racing to build superintelligence before international rivals. Therefore, we aim our research at improving the most likely situations we might find ourselves in and moving the world in safer directions.

\subsection{High-level scenarios for AI development} \label{sec:high_level_scenarios}

\begin{figure}[htbp]
    \centering
    \includegraphics[width=1\textwidth]{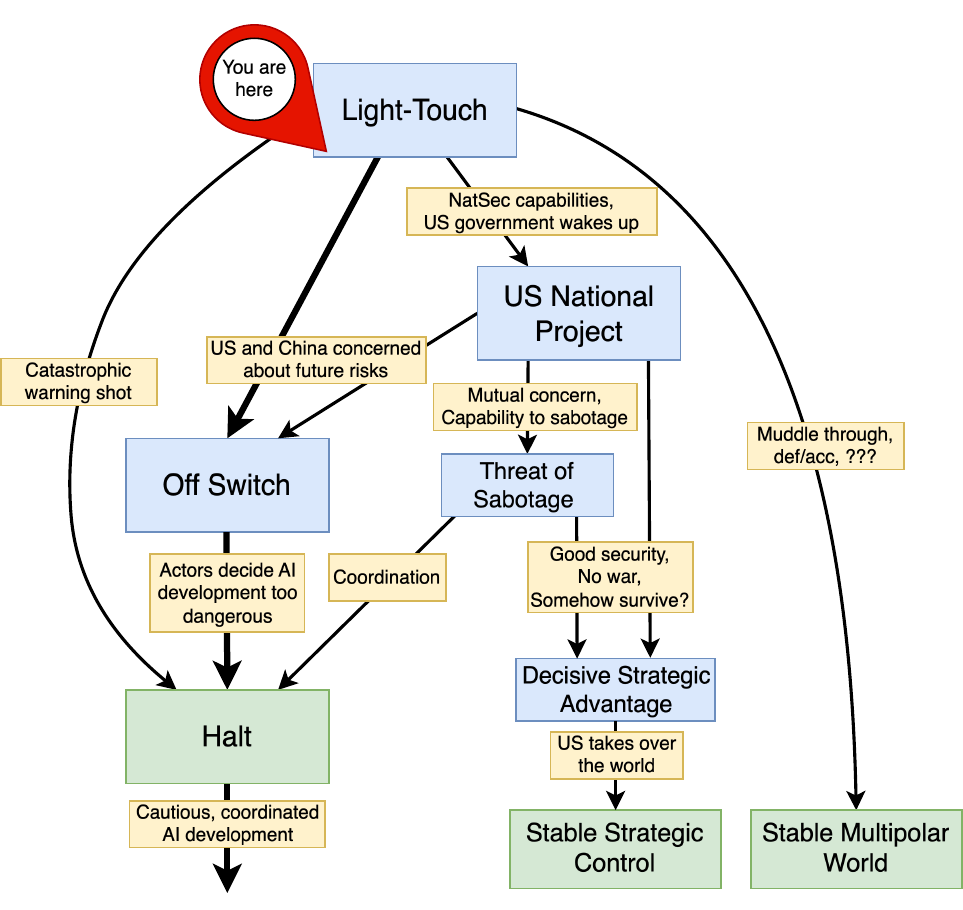} 
    \caption{A rough tree of how AI governance paradigms may evolve, assuming catastrophe is avoided at each step. This is simplified, and one could draw many more connections between the scenarios. This figure is duplicated in the Executive Summary (Figure~\ref{fig:paradigm_tree_exec_summary}).}
    \label{fig:paradigm_tree_intro}
\end{figure}

This section explores four potential trajectories for advanced AI development and governance: a coordinated global Halt on dangerous AI development (\hyperref[sec:halt_off_switch]{Off Switch and Halt}), a US National Project leading to US global dominance (\hyperref[sec:us_national_project]{US National Project}), continued private-sector development with limited government intervention (\hyperref[sec:light_touch]{Light-Touch}), and a world where nations maintain AI capabilities at safe levels through mutual threats of interference (\hyperref[sec:threat_of_sabotage]{Threat of Sabotage}). While we believe the Halt represents the most promising path forward, understanding how these different scenarios might unfold---and how the world could transition between them---is crucial for developing effective governance strategies.

The \textbf{\hyperref[sec:halt_off_switch]{Halt}} is our desired strategy for the future of AI development, where the world coordinates to shut down dangerous AI activities. This Halt could fall out of one of the other scenarios discussed below, or it could be directly pursued from the current situation. For example, world leaders and AI companies could decide they want to build an Off Switch tomorrow (though the process would likely take months or more). A successful Halt would give humanity many years to build our understanding of AI systems, in turn allowing cautious and low-risk development when appropriate. It is an open question how humanity would move forward in AI development once a global Halt is instituted---e.g., via a single collaborative scientific project similar to CERN---but we largely leave such questions to the future. With sufficient time, a cautious AI development project could take place, but the first major hurdle is buying the necessary time.

In the \textbf{\hyperref[sec:us_national_project]{US National Project}} scenario, the US Government assists with security at domestic companies and implements export controls on AI. The government then becomes heavily involved in AI development via an AI “US National Project”. The US National Project has two main goals. First, preventing access to powerful AI by potential adversaries, including other nations (primarily China) and rogue actors (terrorist groups and hostile states). Second, advancing its own AI capabilities. The US National Project would likely pursue the goal of building artificial superintelligence (ASI), AI much smarter than expert humans across all cognitive domains.

One potential path to ASI is via first automating AI R\&D---a feedback loop sometimes referred to as an \textit{intelligence explosion}. If this approach worked and the resulting ASI was, miraculously, aligned to its developer’s intentions, the US Government could use it to achieve a \textit{decisive strategic advantage}---be it technological, military, or economic---over all other nations. In turn, a decisive strategic advantage would allow the US to quash any competing AI projects and potentially force other nations to submit to a US-led international regime. The strategy depends on several highly uncertain conditions: maintaining a substantial lead in AI capabilities, avoiding sabotage or escalation by rivals, preventing proliferation, safely controlling an intelligence explosion, and successfully converting technological superiority into stable international control.

\textbf{We think the US National Project strategy is extremely dangerous and we do not endorse it}. However, this strategy is prominently \href{https://situational-awareness.ai/}{touted by others}~\citep{aschenbrenner_situational_2024}, so we discuss ways to make this very unsafe project marginally safer. While our primary focus is on a government-controlled project, many of the research questions we explore are also relevant to softer forms of government involvement; the core strategic goal of achieving technological superiority is similar even if implementation differs in important respects. Many of the research projects and interventions for the US National Project will also be helpful for the Off Switch and Halt, and \textit{vice versa}. For example, projects for preventing the proliferation of AI technology and projects for understanding the AI capabilities of different actors will be useful to both.

We also briefly discuss two other likely scenarios for AI development: Light-Touch and Threat of Sabotage. We spend less time on these scenarios because neither appears to be sufficiently stable and risk-reducing. However, these are situations that seem somewhat likely, and the interplay between them and an Off Switch or US National Project is useful to consider.

In the current world, governments maintain a \textbf{\hyperref[sec:light_touch]{Light-Touch}} relationship with AI companies, allowing private sector AI development to proceed with minimal intervention. We believe this state is uniquely dangerous for a couple of reasons: the risk of foreign actors stealing AI models or secrets, and race dynamics between AI companies. On the current trajectory, AI companies will be unable to implement state-proof security measures before they develop advanced AI systems, these AI systems will be stolen by US adversaries, and there will be an intense competition involving US companies and foreign actors. Racing through advanced AI development is very dangerous because there are likely to be unforeseen difficulties as AI capabilities advance above human level, so we should give ourselves time to solve these difficulties rather than rushing ahead.

It is plausible that the US Government will be aware of these risks (especially the risks of foreign actors stealing model weights) and intervene, for example by requiring and assisting with state-proof security. But other than this, the government may take a mostly light-touch approach to the leadership and direction of the AI developers, leaving the CEOs or boards in control. While a Light-Touch scenario is likely in the short-term, we do not believe it will be tenable forever, as private companies develop AI systems that significantly affect national security (e.g., ability to create novel WMDs). Advanced AI capabilities are going to be very important on the national and international stage, and it is difficult to tell stories where the government does not become more involved.

\textbf{\hyperref[sec:threat_of_sabotage]{Threat of Sabotage}} is similar to \textit{Mutual Assured AI Malfunction (MAIM)} described in \textit{\href{https://www.nationalsecurity.ai/}{Superintelligence Strategy}}~\citep{hendrycks_superintelligence_2025}. In Threat of Sabotage, great powers (i.e., the US and China) threaten to interfere with each other’s AI development, creating a halt through threats (including, potentially, ones actually carried out) rather than cooperation. In effect, countries would threaten to sabotage AI development they perceive as dangerous, thus keeping AI capabilities below some danger threshold.

Threat of Sabotage relies on states having clear visibility into each other’s AI development and a consistent ability to sabotage dangerous development: both visibility and sabotage capabilities need to consistently be faster than the time needed for a rival to gain a decisive advantage. Achieving enough visibility and sabotage capability may require substantial multi-lateral coordination, eventually looking similar to a global Off Switch, with countries closely monitoring each other’s AI activities and retaining the ability to stop dangerous activities. This dynamic could arise in a Light-Touch or US National Project scenario as US rivals become worried about advanced AI development, assuming the security of AI projects is weak enough to allow these threats.

These four scenarios are useful for orienting to the situation, but they are illustrative in nature. Reality will likely involve some mix of the scenarios and modifications of them; for example, a successful US National Project will involve nonproliferation efforts that are also useful for the Off Switch. The key analysis and many of the research questions also apply to slightly modified versions of these scenarios.  
\\ 
\begin{mdframed}[backgroundcolor=red!20]
\centering
\textbf{We believe the US National Project, Light-Touch, and Threat of Sabotage strategies are reckless and dangerous. We do not endorse them.}
\end{mdframed}

These scenarios can be evaluated and compared based on the core risks of concern. We provide such an evaluation, along with a brief list of pros and cons of each scenario. These assessments are our analysis, mainly following from the general description of the scenarios.

\begin{table}[htbp]
\centering
\caption{The main pros and cons of the scenarios and how much of each core risk they involve. Ratings are based on our analysis of the scenarios. Color coding reflects the risk level. (Same as Table \ref{tab:scenario_ratings_exec_summary})}
\label{tab:scenario_ratings_intro}
\begin{NiceTabular}{@{} >{\raggedright\arraybackslash}p{\ColScenario} 
                   >{\arraybackslash}p{\ColProsCons} 
                   >{\arraybackslash}p{\ColProsCons} 
                   >{\centering\arraybackslash}m{\ColRisk}      
                   >{\centering\arraybackslash}m{\ColRisk}      
                   >{\centering\arraybackslash}m{\ColRisk}      
                   >{\centering\arraybackslash}m{\ColRisk}      
                }[cell-space-limits = 4pt] 
\toprule
Scenario & Pros & Cons & \Block[c]{}{\mbox{Loss of}\\Control} & Misuse & War & \Block[c]{}{Bad\\\mbox{Lock-in}} \\
\hline

Off Switch and Halt &
    \parbox[t]{\hsize}{%
    \raggedright 
    \begin{itemize}[leftmargin=*, label=\textbullet, nosep, topsep=0pt, itemsep=0pt, partopsep=0pt, parsep=0pt] 
        \item Careful AI \mbox{development}
        \item International \mbox{legitimacy}
        \item Slow societal \mbox{disruption}
    \end{itemize}
    } & 
    \parbox[t]{\hsize}{%
    \raggedright 
    \begin{itemize}[leftmargin=*, label=\textbullet, nosep, topsep=0pt, itemsep=0pt, partopsep=0pt, parsep=0pt]
        \item Difficult to implement
        \item Not a complete \mbox{strategy}
        \item Slow AI benefits
    \end{itemize}%
    } &
    \Block[fill=MyGreen]{1-1}{Low} & \Block[fill=MyGreen]{1-1}{Low} & \Block[fill=MyGreen]{1-1}{Low} & \Block[fill=MyMedium]{1-1}{Mid} \\
\hline
US National Project &
    \parbox[t]{\hsize}{%
    \raggedright 
    \begin{itemize}[leftmargin=*, label=\textbullet, nosep, topsep=0pt, itemsep=0pt, partopsep=0pt, parsep=0pt]
        \item Centralized ability to implement safeguards
        \item Limited proliferation
    \end{itemize}%
    } &
    \parbox[t]{\hsize}{%
    \raggedright 
    \begin{itemize}[leftmargin=*, label=\textbullet, nosep, topsep=0pt, itemsep=0pt, partopsep=0pt, parsep=0pt]
        \item Arms race
        \item Breaking international norms
    \end{itemize}%
    } &
    \Block[fill=MyRed]{1-1}{High} & \Block[fill=MyGreen]{1-1}{Low} & \Block[fill=MyRed]{1-1}{High} & \Block[fill=MyRed]{1-1}{High} \\
\hline
Light-Touch &
    \parbox[t]{\hsize}{%
    \raggedright 
    \begin{itemize}[leftmargin=*, label=\textbullet, nosep, topsep=0pt, itemsep=0pt, partopsep=0pt, parsep=0pt]
        \item Fast economic benefit
        \item Less international provocation
        \item Easy to implement (default)
    \end{itemize}%
    } &
    \parbox[t]{\hsize}{%
    \raggedright 
    \begin{itemize}[leftmargin=*, label=\textbullet, nosep, topsep=0pt, itemsep=0pt, partopsep=0pt, parsep=0pt]
        \item Corporate racing
        \item Proliferation
        \item Limited controls \mbox{available}
        \item Untenable
    \end{itemize}%
    } &
    \Block[fill=MyRed]{1-1}{High} & \Block[fill=MyRed]{1-1}{High} & \Block[fill=MyMedium]{1-1}{Mid} & \Block[fill=MyMedium]{1-1}{Mid} \\
\hline
Threat of Sabotage &
    \parbox[t]{\hsize}{%
    \raggedright 
    \begin{itemize}[leftmargin=*, label=\textbullet, nosep, topsep=0pt, itemsep=0pt, partopsep=0pt, parsep=0pt]
        \item Slower AI development
        \item Limited cooperation needed
    \end{itemize}%
    } &
    \parbox[t]{\hsize}{%
    \raggedright 
    \begin{itemize}[leftmargin=*, label=\textbullet, nosep, topsep=0pt, itemsep=0pt, partopsep=0pt, parsep=0pt]
        \item Ambiguous stability
        \item Escalation
    \end{itemize}%
    } &
    \Block[fill=MyMedium]{1-1}{Mid} & \Block[fill=MyGreen]{1-1}{Low} & \Block[fill=MyRed]{1-1}{High} & \Block[fill=MyMedium]{1-1}{Mid} \\
\bottomrule 

\CodeAfter
  \tikz \draw [gray] 
    (2-|4) -- (6-|4)  
    (2-|5) -- (6-|5)  
    (2-|6) -- (6-|6)  
    (2-|7) -- (6-|7)  
    (2-|8) -- (6-|8); 
\end{NiceTabular}
\end{table}

\subsection{Research project taxonomy} \label{sec:research_taxonomy}

We've slightly organized the projects discussed below based on how concrete they are, whether TGT is well placed to work on them ourselves, and whether they should be done in the future rather than now.
\begin{itemize}
    \item \NC{} (not concrete): Items not concrete enough to be specific projects; high-level questions.
    \item \GOV{} (government): Projects that are best suited to the people inside the US Government.
    \item \DEV{} (AI developers): Projects that are best suited to AI developers themselves.
    \item \F{} (future): Not currently tractable but in the future may be. Note that there is a spectrum where early work can be useful. Some of these are things that we expect to change a lot over time.
    \item The \impq{} symbol indicates a project or question we think is especially important. These questions and their sub-questions correspond to about 40\% of projects.
\end{itemize}

Expert understanding of these topics often far exceeds that of policymakers and the public. While not explicitly noted throughout, many of these projects would benefit from clear explanatory materials for general audiences.

Below are the questions and research directions we're most excited about (\impq{} projects) and which scenarios they fit into. These are a subset of the questions in the main text, and many of them have multiple sub-questions and motivating context provided later.

\begin{table}[ht]
\centering
\setlength{\tabcolsep}{5pt}
\renewcommand{\arraystretch}{1.3}
\begin{adjustbox}{max width=\textwidth}

\begin{tabular}{p{0.66\linewidth}|>{\centering\arraybackslash}p{0.068\linewidth}|>{\centering\arraybackslash}p{0.068\linewidth}|>{\centering\arraybackslash}p{0.068\linewidth}|>{\centering\arraybackslash}p{0.068\linewidth}|>{\centering\arraybackslash}p{0.068\linewidth}}

\multicolumn{1}{p{0.66\linewidth}}{ 
    {\fontsize{16}{16}\selectfont Research Questions}
} &
\multicolumn{1}{>{\centering\arraybackslash}p{0.068\linewidth}}{ 
    \rotatebox{45}{\parbox{3.3cm}{\raggedright Off Switch and Halt}}
} &
\multicolumn{1}{>{\centering\arraybackslash}p{0.068\linewidth}}{ 
    \rotatebox{45}{\parbox{3.3cm}{\raggedright US National Project}}
} &
\multicolumn{1}{>{\centering\arraybackslash}p{0.068\linewidth}}{ 
    \rotatebox{45}{\parbox{3cm}{\raggedright Light-Touch}}
} &
\multicolumn{1}{>{\centering\arraybackslash}p{0.068\linewidth}}{ 
    \rotatebox{45}{\parbox{3cm}{\raggedright Threat of Sabotage}}
} &
\multicolumn{1}{>{\centering\arraybackslash}p{0.068\linewidth}}{ 
    \rotatebox{45}{\parbox{2cm}{\raggedright Understanding\\[-0.7ex]\hspace{3ex}the world}}
} \\

\hline
\hyperlink{sec:what_compute_monitored}{What are the trends in compute requirements for frontier AI systems?} & \CIRCLE & \CIRCLE & \LEFTcircle & \LEFTcircle & \CIRCLE \\
\hline
\hyperlink{sec:viability_compute_governance}{How viable is compute governance? Specifically, can the US Government prevent China from gaining sufficient compute to have dangerous models? How can governments use controls on specialized chips to institute a long-term moratorium on dangerous AI?} & \CIRCLE & \CIRCLE & & & \CIRCLE \\
\hline
\hyperlink{sec:off_switch_design}{What is the detailed design of the Off Switch and how does the world build it?} & \CIRCLE & \LEFTcircle & & & \\
\hline
\hyperlink{sec:build_common_understanding}{How do we build common understanding about AI risks and get buy-in from different actors to build the Off Switch?} & \CIRCLE & \LEFTcircle & \LEFTcircle & \LEFTcircle & \LEFTcircle \\
\hline
\hyperlink{sec:emergency_response_plan}{What is a suitable emergency response plan, for both AI projects and the US government? How should actors respond to an AI emergency, including both mitigating immediate harm and learning the right lessons?} & \CIRCLE & \CIRCLE & \CIRCLE & \LEFTcircle & \\
\hline
\hyperlink{sec:how_monitor_compute}{How can governments monitor compute they know about, especially to ensure it isn't being used to violate a Halt?} & \CIRCLE & \LEFTcircle & & \LEFTcircle & \\
\hline
\hyperref[sec:security_algorithmic_secrets]{Security for algorithmic secrets} & \CIRCLE & \CIRCLE & \LEFTcircle & \CIRCLE & \\
\hline
\hyperref[sec:other_levers]{Other than compute and security, what levers exist to control AI development and deployment?} & \CIRCLE & \CIRCLE & & & \LEFTcircle \\
\hline
\hyperlink{sec:understanding_forecasting}{Understanding and forecasting model capabilities} & \LEFTcircle & \LEFTcircle & \LEFTcircle & \LEFTcircle & \CIRCLE \\
\hline
\hyperlink{sec:trends_cost_AI_inference}{What are the trends in the cost of AI inference?} & \LEFTcircle & & & \LEFTcircle & \CIRCLE \\
\hline
\hyperref[sec:measure_us_lead]{How do you measure the US lead?} & \LEFTcircle & \CIRCLE & \LEFTcircle & \LEFTcircle & \LEFTcircle \\
\hline
\hyperlink{sec:bring_in_other_AI_projects}{How could a centralized US National Project bring in other AI development projects (domestic and international)?} & \LEFTcircle & \CIRCLE & & & \\
\hline
\hyperlink{sec:ready_to_go_research}{What ready-to-go research should the US National Project prioritize using AIs for, when AIs are capable of automating AI safety research?} & & \CIRCLE & \LEFTcircle & & \\
\hline
\hyperlink{sec:safety_plan}{What is a safety plan that would allow an AI project to either successfully build aligned advanced AI, or safely notice that its development strategy is too dangerous?} & \LEFTcircle & \CIRCLE & \CIRCLE & & \LEFTcircle \\
\hline
\hyperlink{sec:reduce_racing}{What mechanisms are available to reduce racing between nations?} & \LEFTcircle & \CIRCLE & & \LEFTcircle & \\
\hline
\hyperlink{sec:what_is_the_dsa}{How might the US National Project achieve a decisive strategic advantage using advanced AI?} & & \CIRCLE & & \CIRCLE & \\
\hline
\hyperlink{sec:usnp_realize_too_dangerous}{How does the US National Project realize that its strategy is too dangerous?} & & \CIRCLE & & & \LEFTcircle \\
\hline
\hyperlink{sec:interventions_coordinate_domestic}{What interventions are available to coordinate the domestic AI projects to reduce corporate race dynamics, while US Government intervention is light-touch?} & \LEFTcircle & & \CIRCLE & & \\
\midrule
\end{tabular}
\end{adjustbox}
\caption{The most important research questions or projects and which scenario they pertain to. Questions which are very helpful for a scenario are marked \CIRCLE, questions which are medium helpful are marked \LEFTcircle.}
\end{table}

\clearpage
\section{Off Switch and Halt} \label{sec:halt_off_switch} 

There are many situations where the world might desire to shut down dangerous AI activities (both development and deployment). We believe a global Halt could be critical to reduce extinction risk from advanced AI systems, enabling humanity to first build its scientific understanding of AI systems before bringing such systems into existence. This Halt needs to be global, needs to cover all forms of dangerous AI activities (i.e., potentially covering pre-training, post-training, and deployment), and needs to last for years or longer. The rest of the world is not yet in agreement that a Halt is necessary. Therefore, our main ask is that humanity \textbf{preserve the option to Halt}.

We conceptualize this optionality as an \textit{Off Switch} for AI. An Off Switch includes the technical, legal, and institutional infrastructure needed for a legitimate authority (e.g., the US Government, the UN Security Council, or some coalition of nation-states) to globally halt all dangerous AI development and deployment for an extended period of time (e.g., decades). Off Switch infrastructure should be developed in advance of when it is needed. Importantly, there may be levels of an Off Switch, for instance, different lengths of a halt or different types of AI activities that must be halted. These levels may allow for more targeted use of an Off Switch, thus making Off Switch infrastructure more useful to mitigate various threats; mitigating risks from human misuse could look different from mitigating risks from rogue AI systems.

Building an Off Switch involves implementing the technical means to rapidly identify and shut down unsafe AI systems and AI projects, as well as the capacity to ensure that there is no dangerous AI development, globally.

There are no perfect historical analogies for this Off Switch and Halt. Analogous cases often involve technology with modest short-term benefits, whereas AI is likely to be extremely useful (economically and militarily) in advance of when progress must stop. The AI case may also require more substantial monitoring than other technologies; this is because the infrastructure required for dangerous AI activities has many other uses, and it may be challenging to differentiate dangerous from benign AI activities. Despite these disanalogies, there are some related examples. The Biological Weapons Convention bans biological weapons, including their development, production, and use. In another case of banning dangerous technology, the 1987 Montreal Protocol phased out the use of ozone depleting substances. While the precedent is not as strong as we might hope, the world has sometimes coordinated to avoid collectively destructive technology.

In response to this proposal of a Halt, some may be worried about the risk of authoritarianism enabled by unilateral control over advanced AI. Among methods of implementing a global Halt, a coordinated Off Switch is the approach that appears to run the least risk of enabling authoritarianism. In particular, if a global Halt happens as a result of the US National Project achieving a decisive strategic advantage, the power exercised over the rest of the world would be very unilateral. By contrast, we want the international community to come together early to build an Off Switch which is collectively enforced and is unlikely to be hijacked for authoritarian ends. In particular, this Off Switch could involve a multilateral treaty with symmetric compliance requirements. We don’t see any credible, safe paths through the development of superintelligence that do not involve some degree of coordinated control over AI development. Principally, this is because such AI systems will be incredibly dangerous and we do not expect defensive technology to always be sufficient to prevent harm. We think building an Off Switch and preparing to Halt global AI development is an approach that runs relatively little risk of authoritarian control.

This section discusses what research might assist in building an Off Switch ahead of time and what might be needed for a successful Halt. First, we discuss foundational work to conceptualize the Off Switch and determine how it could be constructed. An Off Switch is our desired path to a global Halt, including coordination and careful preparation. However, pressing the Off Switch is only one path to a Halt. A global Halt might also emerge through less structured paths, such as a US National Project quickly pivoting to implement a Halt if continued development appears too dangerous.

We then discuss three stages of a Halt, which are applicable regardless of how humanity arrives at this Halt strategy. The first of these, \hyperref[sec:acute_risk]{Acute risk mitigation}, involves preparing for emergencies and how to respond to them. For example, one such emergency could be a misaligned AI self-exfiltrating from its developers and attempting to gain power. Emergencies could serve as a \textit{warning shot} and spur serious efforts to Halt, but only if humanity survives them. This acute risk mitigation stage is not strictly necessary because humanity might instead choose to halt dangerous AI development before facing imminent catastrophe. The next stage is a \hyperref[sec:short_term_pause]{Short-term pause} (lasting months to years), buying time to develop longer term solutions. The third stage is a \hyperref[sec:long_term_moratorium]{Long-term moratorium} (lasting years or longer), creating a sustainable path forward for humanity’s AI development.

\subsection{Off Switch} \label{sec:off_switch}

\begin{itemize}
    \item \hypertarget{sec:off_switch_design}{} \impq{} What is the detailed design of the Off Switch, and how does the world build it? This is particularly important because Off Switch plans currently don't have a concrete goal. Laying out a detailed design will enable other work toward useful intermediate goals in building an Off Switch. 
        \begin{itemize}
            \item What are reasonable levels of pressing an Off Switch based on different levels of political will?
            \item How could an Off Switch interact with and slot into the existing international political climate?
            \item What is the reasonable path to building an Off Switch? What is one concrete story of what this could look like, escalating through levels of Off Switch?
            \item What is the minimum viable product (MVP) of an Off Switch? For example, an Off Switch operating only at the US domestic level without international coordination.
            \item If the US president wanted to halt AI development tomorrow, could they do so? What legal powers would be relevant? How could they influence foreign AI development?
            \item Which Off Switch components need to be developed significantly in advance, and which could be implemented with limited notice?
            \item \NC{}: Scenario planning, war-gaming, and similar approaches may be useful for better clarifying what an Off Switch might look like and other details around it.
        \end{itemize}
    \item Institutionally, what might an Off Switch look like, within the US or internationally?
        \begin{itemize}
            \item Which authorities/parties are best placed to have control over an Off Switch?
            \item What would a clear chain of command look like for shutting down domestic AI training or deployment?
            \item The Off Switch needs to be utilized at the right time to avoid danger, and false alarms could be costly. What is the right balance of false positives to false negatives? How could this balance be achieved?
            \item Which government agencies are currently well suited for being in charge of the domestic Off Switch?
            \item What new institutions might be needed to be in charge of an international Off Switch (e.g., an IAEA for AI)?
        \end{itemize}
    \item\hypertarget{sec:off_switch_proliferation}{} What level of AI proliferation would render an Off Switch infeasible? 
        \begin{itemize}
            \item What capability evaluation should be used to decide if a model should not be proliferated? Examples: 5x AI R\&D uplift, OpenAI Preparedness Critical, “the world could not implement an Off Switch if this model were open sourced and proliferated”.
            \item To what extent do open-source models---or private models in non-frontier projects---have concerning capabilities? Where are capabilities likely to be in the near future?
                \begin{itemize}
                    \item[\linkbullet] See \hyperlink{sec:understanding_forecasting}{Understanding and forecasting model capabilities}.
                \end{itemize}
            \item What are the avenues by which the government could prevent the open-sourcing of models that pose proliferation risk or are near posing proliferation risk?
            \item For different classes of actors, including states and terrorists, which AI capabilities pose significant risks? Example: Some AI capabilities, like PhD-level biological weapon development uplift, may elevate the capabilities of non-state actors but not differentially affect those of states.
            \item Are there ways to delay the deployment of autonomous AI systems which might themselves be ungovernable by an Off Switch? For example, an international agreement around monitoring requirements for autonomous AI systems.
        \end{itemize}
    \item When should an Off Switch be pressed? An Off Switch would likely have different levels of intervention, and it is important to determine when it is best to activate each of them.
        \begin{itemize}
            \item Conceptual work to understand which AI capabilities, traits, and behaviors are most concerning
                \begin{itemize}
                    \item[\refbullet{}] \cite{karnofsky2024sketch, raman2025intolerable, metr2025common, shevlane_model_2023}
                \end{itemize}
            \item Conceptual work to understand what might be “points of no return”
                \begin{itemize}
                    \item[\refbullet{}] \cite{kokotajlo_date_2020}
                \end{itemize}
            \item Empirical evaluations and continuous monitoring to understand whether AI systems are at concerning levels
                \begin{itemize}
                    \item[\refbullet{}] \cite{noauthor_helm_nodate, liang2022holistic, wijk_re-bench_2024}
                \end{itemize}
            \item Forecasting work to predict the levels of those capabilities in the coming months or years. These questions are especially difficult because they must be forward-looking---that is, if the Off Switch is pressed too late, humanity loses.
                \begin{itemize}
                    \item[\linkbullet] See \hyperlink{sec:understanding_forecasting}{Understanding and forecasting model capabilities}.
                \end{itemize}
            \item Consensus building about risk levels. For example, develop international red lines.
                \begin{itemize}
                    \item[\refbullet{}] \cite{idais_international_2024}
                \end{itemize}
        \end{itemize}
    \item \hypertarget{sec:build_common_understanding}{} \impq{} How do we build common understanding about AI risks and get buy-in from different actors to build the Off Switch? This is very important because broad buy-in from key decision makers is the most difficult hurdle to building an Off Switch. 
        \begin{itemize}
            \item How do we convince and show relevant decision makers that an Off Switch is a good idea? Examples: communicating concerns about misuse, misalignment, and power disruption.
            \item What are the main objections decision makers will have to building the Off Switch? Some of these will be legitimate and may require changing the plan, while some will require better communication.
            \item What is the best language to use when communicating about these topics? What are terms to use or avoid? What will be clear to various audiences?
            \item Build harmless demonstrations of dangerous model capabilities, write good explanations of AI risks, communicate with the public and key decision makers.
            \item Create and maintain a database of bad behaviors and misalignment-related failures in production AI systems.
                \begin{itemize}
                    \item[\refbullet{}] \cite{krakovna_specification_2018, deepmind_goal_2022}
                \end{itemize}
        \end{itemize}
    \item How could an international agreement move the world toward having an Off Switch? What countries could be involved? What could the agreement aim to accomplish?
    \item Off Switch security
        \begin{itemize}
            \item Authority and institutional security: How can Off Switch infrastructure be secured to avoid illegitimate power grabs or use by unauthorized authorities?
            \begin{itemize}
                \item[\refbullet{}] \cite{davidson2025aienabledcoupsho} 
            \end{itemize}
            \item Technical security: How can Off Switch infrastructure be hardened to prevent security breaches, such as a misaligned AI shutting off other AI activities, or human hacking?
        \end{itemize}
\end{itemize}

\subsection{Acute risk mitigation} \label{sec:acute_risk} 

The first stage of the Halt may be \textbf{acute risk mitigation}. For example, the Halt may be triggered by evidence that an AI system poses an acute risk of catastrophic harm. If this is the case, the top immediate priority will be to reduce this risk. We lay out questions about what warning shots we might expect, and how we should prepare for them.

\begin{itemize}
    \item \hypertarget{sec:warning_shots}{} What things might serve as warning shots? Examples: autonomous AI slowly accruing resources, small scale disaster, highly capable AIs, AIs capable in national security-relevant domains, AI attempting but failing at some misaligned action. 
        \begin{itemize}
            \item Empirically, evidence for AI risks is interpreted differently by different people. Evidence that some people find convincing is sometimes not compelling at all to others. What warning shots are likely to be important to different key decision makers? Based on this, which warning shot-related projects should researchers pursue?
        \end{itemize}
    \item What can an AI project do if they catch the AIs scheming or “red-handed”? How could this serve as a warning shot? How should the AI project, US Government, and wider world respond to this particular warning shot?
        \begin{itemize}
            \item[\refbullet{}] \cite{greenblatt_catching_2024,ryan_greenblatt_how_2025, shlegeris_would_2024}
        \end{itemize}
    \item What early warning systems could enable rapid response to acute risks? How can the relevant institutional capacity be created in government and AI developers to correctly understand and respond to these risks? What should these systems look like at a domestic and international level?
    \item How can we proactively set up controlled scenarios that serve as convincing warning shots, and are not dangerous? How do we ensure that the first warning shot isn’t caused by AI systems that are uncontrollable and precipitate human extinction? Examples: model organisms, forecasting, defense in depth to catch exfiltration attempts.
    \item \hypertarget{sec:emergency_response_plan}{} \impq{} What is a suitable emergency response plan, for both AI projects and the US government? How should actors respond to an AI emergency (i.e., a warning shot), including both mitigating immediate harm and learning the right lessons? This is particularly important because humanity needs to both survive a specific emergency and then react appropriately, changing how we govern advanced AI after the emergency has ended and installing robust solutions rather than shallow, local patches. 
        \begin{itemize}
            \item How would the world go about shutting down a highly capable autonomous AI system that has self-exfiltrated? There may be analogies to computer worms; potentially useful interventions could include Know Your Customer (KYC) rules, watermarking, and AI text detection.
            \item Are there low-cost interventions that could prepare the world in advance? Examples: broadly used and coordinated KYC, proof-of-humanity for some tasks.
            \item Historical review of response to emergencies and catastrophes.
            \item Review the existing emergency response plans in other fields.
            \item \GOV{}: How do we stop cascading failures \textit{in the world} via fail-safe measures and circuit breakers? Examples: stock market, power grid, software systems. Actors within the US Government may be best positioned for these projects because they will likely require access and influence over important societal systems, which civilian actors usually will not have. Non-government actors could make progress here by prototyping fail-safe measures.
            \item How do we stop cascading failures \textit{in AI systems}? These failures may look like jailbreaks or coordinated defection by misaligned AI systems. What can be done to prevent these failures?
                \begin{itemize}
                    \item[\refbullet{}] \cite{scher_idea_2024, greenblatt_ai_2024, cohen_here_2025}
                \end{itemize}
        \end{itemize}
    \item What can be done to ensure the long-term response to acute risk robustly addresses the whole risk landscape? How can a warning shot translate into useful changes?
        \begin{itemize}
            \item Historical case study of things like warning shots and the reaction to them---what leads to effective change?
                \begin{itemize}
                    \item[\refbullet{}] \cite{guest_prospects_2023}
                \end{itemize}
            \item How could existing institutions and frameworks, such as insurance and liability, play a role in reducing AI risks? How should these be changed to better apply to the AI case?
            \item What incident reporting systems could allow for proper sharing of critical information after risks have been mitigated? How can the relevant institutional capacity be created in government and AI developers to correctly understand and respond to these risks? What should these systems look like at a domestic and international level?
        \end{itemize}
    \item What are the key defenses to prevent harm, if a model has exfiltrated? In the future, there may be autonomous AI systems which appear threatening but are not imminently causing harm. It could be important to harden the world against them in various ways. Examples: pathogen screening, international norms around not cooperating with misaligned AGIs, epistemic defenses.
        \begin{itemize}
            \item What are the main threats such a system might pose? E.g., ARA threat modeling~\citep{the-rogue-replication-threat-model}.
            \item[\linkbullet] See \hyperref[sec:defensive_acceleration]{Defensive acceleration}.
        \end{itemize}
    \item What will be the critical infrastructure to defend, and what will be needed to defend it, assuming a model has exfiltrated? Example: securing human-led AI projects against biological weapon attacks.
    \item What redundancy exists to implement emergency response protocols? Example: if a data center needs to be shut off, are there multiple independent approaches to doing so?
    \item Preventing harm from advanced AI systems may require multilateral defense efforts. How can top cyber-capable and defense institutions collaborate to respond to AI emergencies and prevent large-scale harm? Example: joint cyber-military exercises.
\end{itemize}

\subsection{Short-term pause} \label{sec:short_term_pause} 

When a Halt is initiated, it needs to impose a global moratorium on dangerous AI development and deployment. Governments could use control over AI compute or other levers to ensure dangerous AI activities do not take place. If the US has a large enough lead, it may be sufficient in the short term for the US Government to implement a pause domestically. But if the US lead is small or non-existent, then the pause will have to include China as well. Eventually, the world will pivot into a longer-term moratorium intended to last years or decades, but the infrastructure for a long-term Halt will not be ready immediately. Therefore, it is worth discussing mechanisms that enable a short-term pause, buying time for a more sustainable state to be reached.

\subsubsection{To what extent does a Halt need to be international, and on what time scale?}

Due to difficulties with international coordination, it may initially be easiest for the US to institute a domestic Halt on AI development. However, this approach is unlikely to be stable in the long term. Other nations would eventually advance their own AI capabilities, potentially creating systems that pose significant misalignment risks and threaten American sovereignty. One strategy could involve the US first implementing a domestic Halt, then leveraging the existing lead time to negotiate a comprehensive, internationally coordinated Halt. This domestic halt would both buy time for an international agreement and provide a clear and costly signal that the US takes risks from advanced AI seriously, potentially making international agreements more likely. Alternatively, it may be more politically viable for the US to force other countries to halt first, or for all countries to pause simultaneously.

\begin{itemize}
    \item Should the short-term pause be domestic first and then international, everybody at once, non-US first, or something else? The political and operational viability of these different approaches may affect what can happen.
    \item How much time does the short-term pause (possibly domestic US) need to buy to develop other solutions?
    \item \hypertarget{sec:us_lead_over_china}{} \impq{} How much of a lead time do American companies have over Chinese companies in AI development? This is important for the Halt strategy because, if the US has a significant lead, US projects may be able to pause and use their lead time to develop other solutions (e.g., verification mechanisms, diplomacy). If US projects do not have a significant lead, they would likely need to coordinate a simultaneous pause with China. 
    \begin{itemize}
        \item[\refbullet{}] See \hyperref[sec:measure_us_lead]{How do you measure the US lead?}
    \end{itemize}
    \item How likely are other countries (not US or China) to be competitive in frontier AI, for example due to large domestic compute capacity or obtaining model weights from US developers?
    \item Many frontier US AI companies use international subsidiaries and contractors to bypass regulations and taxes~\citep{setser_spotty_2024}. Will these complex international corporate structures cause problems for the US Government’s ability to shut down dangerous AI development?
    \item What can be done to increase US centralization of AI development? Examples: build frontier AI data centers in the US and allied democracies; ensure AI chip export controls remain tight; subsidize US AI development. Some interventions may accelerate AI development and hence pose externalities; which interventions increase US centralization without substantially increasing risk?
\end{itemize}

\subsubsection{International agreements}

International agreements will be essential during the short-term pause phase of a Halt. We need to clarify precisely what activities are prohibited (or permitted, if a cautious approach is adopted), establish effective methods for monitoring and verification, and determine clear responses to violations.

\begin{itemize}
    \item What international agreements could enable and uphold a short-term pause?
        \begin{itemize}
            \item What are the plausible components that such an agreement could include? Examples: verification measures, list of prohibited activities, enforcement mechanisms.
            \begin{itemize}
                    \item[\refbullet{}] \cite{scholefield2025international}
                \end{itemize}
            \item What existing international institutions could oversee the implementation and enforcement of an international AI Halt agreement?
        \end{itemize}
    \item What does the escalation ladder look like?
        \begin{itemize}
            \item \GOV{}: How should governments respond to potential violations? This question is best suited for the US Government because the relevant expertise is largely housed there.
            \item What are realistic ways of doing secondary verification---actions that are more costly and more reliable, and used if the first monitors trigger? Examples: sending in inspectors, going through the logs.
                \begin{itemize}
                    \item[\refbullet{}] \cite{scher_mechanisms_2024}
                \end{itemize}
            \item \GOV{}: What are the back-channels for diffusing the situation and regaining stability? Who are the key players internationally who would be involved in this response? How can the relevant relationships be built now? Actors within the US Government with access to back-channels and more context on which back-channels exist are likely best positioned to work on this question.
            \item What have escalation and diffusion protocols looked like historically?
        \end{itemize}
    \item How can countries be brought into international agreements about AI? This likely involves a combination of carrots, sticks, and building consensus about risks.
        \begin{itemize}
            \item[\linkbullet] See \hyperlink{sec:build_common_understanding}{How do we build common understanding about AI risks and get buy-in from different actors to build the Off Switch?}
            \item[\linkbullet] See \hyperlink{sec:benefit_sharing}{How can we enable global benefit sharing?}
        \end{itemize}
\end{itemize}

\subsubsection{Leveraging compute to control AI development and deployment}

A key lever for governing AI development is controlling access to the advanced computer chips (sometimes called \textit{compute}) used for AI workloads. Compute is a convenient node for governance because large numbers of chips are detectable, many chips are necessary for advanced AI training, chips are easily quantifiable, and the AI chip supply chain is concentrated~\citep{sastry_computing_2024}. Monitoring compute---such as a chip’s location and usage---can help governments ensure no dangerous AI activities occur.

However, algorithmic progress may significantly reduce the compute required for dangerous AI, and untracked or distributed compute poses additional challenges. Moreover, countries could eventually establish their own domestic chip production, removing the current opportunity for control via the centralized supply chain. So, while compute is currently a useful node for governing advanced AI, this is not guaranteed in the future.

\begin{itemize}
    \item Where is the compute? If we wanted to physically monitor it right now, what would that look like?
        \begin{itemize}
            \item Where are the largest AI data centers?
                \begin{itemize}
                    \item[\refbullet{}] \cite{fist_how_2024, pilz_compute_2023, pilz2025trends}
                \end{itemize}
            \item How many AI chips could not be quickly located and brought under a monitoring regime? What are the timeframes for different monitoring regimes?
            \item AI compute may be co-located with non-AI compute or in sensitive military facilities, which would pose issues for international monitoring. What can be done to mitigate this issue or disincentivize this co-location?
        \end{itemize}
    \item \hypertarget{sec:what_compute_monitored}{} \impq{} What compute needs to be monitored after a Halt is initiated? More generally, what are the trends in compute requirements for frontier AI systems? These questions are important because compute will need to be monitored and restricted during a Halt, but we want to monitor the minimum amount of compute necessary to avoid risks. Trends in compute requirements may mean that over time more compute needs to be monitored, as less compute is required for dangerous AI activities. 
        \begin{itemize}
            \item What scale of harm should be the target of this governance regime and thus the benchmark for dangerous AI activities? For instance, should this regime focus only on extinction-level risks, risks on the order of billions of deaths, millions of deaths, or some other scale.
            \item What constitutes “dangerous AI activities” at this point in time? Does this need to include pre-training, post-training, inference, or other activities? This affects what amount of compute needs to be monitored because different amounts of compute are needed for different dangerous activities. For example, if one could fine-tune a model to make it dangerous, then the threshold for dangerous amounts of compute will be low.
                \begin{itemize}
                    \item How much can AI capabilities or other key traits be improved with post-training? Can post-training make a model more capable in a dangerous way or instill more dangerous traits, such as assisting with harmful requests?
                        \begin{itemize}
                            \item[\refbullet{}] \cite{davidson_ai_2023, measuring-the-impact-of-post-training-enhancements, bowen_data_2024, volkov_badllama_2024}
                        \end{itemize}
                    \item How much can AI capabilities or other key traits be improved with significant inference compute? Can increased inference compute make a model more capable in a dangerous way or instill more dangerous traits?
                        \begin{itemize}
                            \item[\refbullet{}] \cite{deepseek-ai_deepseek-r1_2025, brown_large_2024, epoch2023tradingoffcomputeintrainingandinference, anil_many-shot_2024}
                        \end{itemize}
                    \item What plausible developments in the AI paradigm would substantially affect what compute needs to be monitored? For example, if online/continual learning is successful, monitoring inference may be more important.
                \end{itemize}
            \item Algorithmic progress
                \begin{itemize}
                    \item What is the state of algorithmic progress, and what does it imply about future AI development?
                        \begin{itemize}
                            \item[\refbullet{}] \cite{ho_algorithmic_2024}
                        \end{itemize}
                    \item How much does algorithmic progress depend on compute access for running experiments? Does blocking compute access lead to an increase in algorithmic progress?
                    \item How important are algorithmic secrets likely to be, based on the past and expectations for the future? For example, they may be very important during a \textit{software only intelligence explosion}~\citep{davidson2025threetypesofinte} (rapid advancement in AI capabilities based primarily on AI automation of machine learning research, without significant automation of chip design or chip production).
                        \begin{itemize}
                            \item[\refbullet{}] \cite{erdil_estimating_2024, davidson_what_2023}
                        \end{itemize}
                    \item \DEV{}: What are the most important algorithmic secrets that AI projects should tightly secure? These might be measured by compute multiplier. AI companies are probably best placed to work on this due to their access to algorithmic secrets, but external actors can probably do early work here and may have better incentives for unbiased work.
                        \begin{itemize}
                            \item[\refbullet{}] \cite{davidson_ai_2023}
                        \end{itemize}
                    \item What are the algorithmic progress trends in post-training?
                    \item What are the trends in the cost of AI inference and what might they imply about automated AI R\&D or other notable future developments?
                        \begin{itemize}
                            \item[\linkbullet] See \hyperlink{sec:trends_cost_AI_inference}{What are the trends in the cost of AI inference?}
                        \end{itemize}
                    \item What are the compute requirements for different AI activities? For example, how do the compute requirements compare between pre-training, RL post-training, and inference.
                    \item How will synthetic data and data quality affect model capabilities and scaling?
                \end{itemize}
            \item How should scaling laws, compute scaling, and forecasts of downstream capabilities affect what compute must be monitored?
                \begin{itemize}
                    \item[\linkbullet] See \hyperlink{sec:understanding_forecasting}{Understanding and forecasting model capabilities}.
                \end{itemize}
            \item Take the answers to the previous questions and compile forecasts for trends in compute use and AI capabilities. What do these forecasts imply about export controls, algorithmic secret security, etc.?
                \begin{itemize}
                    \item[\refbullet{}] \cite{kokotajlo2025ai2027} 
                \end{itemize}
            \item What is the state of distributed training, and how does this affect what needs to be monitored?
            \item Are there substantial differences between AI-specific and non-AI compute at the data center level? For example, satellites might not give enough granularity.
                \begin{itemize}
                    \item[\refbullet{}] \cite{heim_limitations_2024}
                \end{itemize}
            \item How likely is non-AI-specific compute to be relevant to a Halt? Examples: gaming GPUs, data center CPUs, consumer CPUs. What is the risk that this compute could be repurposed for dangerous AI development?
                \begin{itemize}
                    \item What would it look like to implement an Off Switch in a world where consumer hardware can be used for dangerous AI activities?
                    \item What fraction of the world’s compute (e.g., by FLOP/s) is made up of these different types of compute?
                    \item[\refbullet{}] \cite{grunewald_are_2024}
                \end{itemize}
            \item Based on the questions above, what compute clusters need to be monitored? This could be determined based on total FLOP/s, power draw, etc.
        \end{itemize}
    \item \hypertarget{sec:how_monitor_compute}{} \impq{} How can governments monitor compute they know about, especially to ensure it isn’t being used to violate a Halt? This is ideally done in a privacy-preserving or zero-knowledge manner. Much of the \href{https://techgov.intelligence.org/research/mechanisms-to-verify-international-agreements-about-ai-development}{verification work} is relevant here~\citep{scher_mechanisms_2024}. This is important because the Halt strategy relies on governments having methods to monitor certain classes of compute and ensuring that compute is not being used for illicit activities. Due to worries about privacy and authoritarianism, compute monitoring will likely be more politically feasible if done using minimally invasive methods. 
        \begin{itemize}
            \item What tools are currently available for compute monitoring? How robust are they?
            \item What tools could be made available in the future for compute monitoring, such as \href{https://yoshuabengio.org/wp-content/uploads/2024/09/FlexHEG-Interim-Report_2024.pdf}{FlexHEGs}~\citep{petrie2024interim}?
                \begin{itemize}
                    \item What needs to change for this to become a reality? What features of current AI chips might enable on-chip governance?
                \end{itemize}
            \item What is the minimum viable product (MVP) for knowing that a large compute cluster is not being used for dangerous AI development in a low-access setting? Examples: power monitoring, interconnect restrictions.
            \item What is the MVP for monitoring workloads in a high-access and medium-trust environment (such as for domestic monitoring)? Build the classifier that cloud providers could use to ensure their customers are following various rules~\citep{heim_governing_2024}; potentially assume access to code.
            \item At different layers of the compute stack, what are the meaningful and difficult-to-spoof differences between different AI workloads? That is, how could workload classification be done at different layers?
            \item Can current inference-optimized chips be viably used for training? How could future chips be made more specifically for one of these purposes? Examples: \href{https://groq.com/about-us/}{Groq}, \href{https://www.etched.com/announcing-etched}{Sohu}, \href{https://cerebras.ai/product-chip/}{Cerebras}, \href{https://aws.amazon.com/ai/machine-learning/inferentia/}{AWS Inferentia}.
                \begin{itemize}
                    \item[\refbullet{}] \cite{rand_deep_2023}
                \end{itemize}
            \item What are the levels of regulation that should be applied to different AI activities and compute? For instance, categories might be: don’t monitor, monitor but don’t restrict, monitor with some restrictions, prohibit.
            \item Draft an initial agreement that countries could use to implement such monitoring. What are the key institutions? When does verification and monitoring take place? What are the key mechanisms to do verification/monitoring, and how are they implemented?
            \item How can a government coalition bring more compute into a central governance framework? What carrots and sticks can be used to incentivize centralization at the domestic and international levels? Examples: benefit sharing, API access, sanctions, ongoing support from certain companies, tax breaks, reducing regulatory burden, compulsion under the DPA.
            \item What is the cost-benefit analysis of monitoring various compute and various AI activities? Should cost-benefit be the primary method for making the determination about what to allow or what to monitor?
            \item[\refbullet{}] \cite{scher_mechanisms_2024, petrie2024interim}
        \end{itemize}
    \item How can we ensure monitored compute remains monitored?
        \begin{itemize}
            \item Is it necessary to centralize compute clusters domestically or internationally in order to reduce the risk of chip smuggling or illicit compute use? Security requirements for data centers to prevent insider (national) threats might mean chips need to be centralized to a few locations.
            \item How can countries verify the presence and completeness of declared compute clusters? Examples: physical inspections, continuous monitoring, portal monitoring, on-chip location tracking.
                \begin{itemize}
                    \item[\refbullet{}] \cite{scher_mechanisms_2024}
                \end{itemize}
            \item What tamper-evident or tamper-proofing mechanisms can be instituted on current or future AI chips to increase the robustness of monitoring?
                \begin{itemize}
                    \item[\refbullet{}] \cite{noauthor_tampersec_nodate}
                \end{itemize}
        \end{itemize}
    \item \hypertarget{sec:compute_breakout}{} Compute breakout time: Another key issue with monitoring compute is the risk that countries might break international agreements and conduct secret AI development or secret AI chip production to support that development. Breakout time refers to the length of time it would take an actor to do particular prohibited AI activities. 
        \begin{itemize}
            \item What are the key bottlenecks along the AI chip supply chain, ideally those specific to AI? How do we ensure these bottlenecks remain relevant?
                \begin{itemize}
                    \item[\refbullet{}] \cite{khan2021semiconductor, grunewald_introduction_2023, thadani2023mapping}
                \end{itemize}
            \item What is the breakout time for doing secret AI chip development given various initial conditions? For instance, how long would it take China or North Korea to produce an H100-class chip, given different realistic states of their chip supply chains?
            \item What is the breakout time on states building covert AI data centers under various conditions: given current data center capacity, given different hardware-enabled governance mechanisms (HEMs)~\citep{kulp2024hardwareenabled}, etc.? This question matters a lot for \hyperref[sec:threat_of_sabotage]{Threat of Sabotage} scenarios where countries need high visibility into AI infrastructure.
            \item Where are the current high-end chip fabrication plants?
            \item \GOV{}, \F{}: If on-chip mechanisms are desired in the future, how can an international coalition verify that these mechanisms are implemented on chips without backdoors or other vulnerabilities? When it comes to actually implementing these mechanisms, all relevant parties will want to ensure that the chips have not been \textit{backdoored}. Implementing and testing these mechanisms will require government involvement.
            \item What is the potential to use lower-end fabs for AI chip production? For instance, is this a 1.3x performance hit, or a 5x performance hit?
                \begin{itemize}
                    \item[\refbullet{}] \cite{grunewald_are_2024, erdil_introducing_2024}
                \end{itemize}
            \item How much could rogue actors scale AI chip production (in violation of a Halt) if this were a very high priority? It’s rumored that one key bottleneck to scaling is TSMC being risk averse and not wanting to lay their own money on the line; how could chip production change if they were given a \$1T government contract?
                \begin{itemize}
                    \item[\refbullet{}] \cite{heim_what_2024}
                \end{itemize}
        \end{itemize}
    \item How can governments prepare in advance for an effective compute governance regime?
        \begin{itemize}
            \item[\linkbullet] See \hyperref[sec:export_controls]{Export controls}.
        \end{itemize}
    \item[\linkbullet] See \hyperlink{sec:viability_compute_governance}{How viable is compute governance?}
\end{itemize}

\subsubsection{Security for AI model weights} \label{sec:security_ai_model_weights}

The theft of model weights would significantly increase proliferation risks and complicate global coordination. AI projects need robust security measures to prevent theft by nation-state adversaries as well as non-state actors. Currently, AI projects fall far short of state-proof security.

\begin{itemize}
    \item[\refbullet{}] \cite{nevo_securing_2024}
    \item \GOV{}, \DEV{}: How do AI projects achieve \href{https://www.rand.org/pubs/research_reports/RRA2849-1.html}{SL5} (security resistant to top state efforts)? Actors within the US Government or AI companies are best positioned to work on this question because they could have more experience dealing with state-level adversaries or direct experience with securing AI model weights.
        \begin{itemize}
            \item How long will it take to achieve this, and where is additional effort most needed in advance?
        \end{itemize}
    \item \NC{}, \GOV{}, \DEV{}: There are many other useful security projects, ranging from blue-sky research to AI developers implementing known security measures. Implementation of tight security may require government support.
    \item What is the likelihood of frontier AI model weights being stolen by adversaries?
    \item How can AI developers achieve the desired level of personnel security, especially given the substantial international ties of many AI company employees?
    \item Is security for AI development bad, due to restricting international visibility? Which security might be bad?
        \begin{itemize}
            \item[\linkbullet] See \hyperlink{sec:what_is_right_level_security}{What is the right level of security for AI projects to have?}
        \end{itemize}
\end{itemize}

\subsubsection[Security for algorithmic secrets]{\impq{} Security for algorithmic secrets} \label{sec:security_algorithmic_secrets}

Algorithmic secrets could significantly impact AI development by enabling powerful models to be trained with fewer resources or by unlocking novel capabilities. Increasingly, these insights remain unpublished due to commercial interests. To prevent misuse by malicious actors, governments and AI developers may need to restrict what information can be published, especially high-risk algorithmic secrets. In addition to legal and institutional changes, strong security measures are also needed to prevent theft, an area likely more neglected than security for AI model weights.

\begin{itemize}
    \item Review historical precedent for information controls around key national security information, identify key analogues and lessons. Examples: how was information security done in the Manhattan Project, how are ex-spies prevented from defecting?
    \item What is the legal precedent and institutional capacity relevant to securing algorithmic secrets? Examples of precedent may include: handling of classified information, Invention Secrecy Act, born secret doctrine, prior restraint (pre-publication censorship), export controls on weapons technology.
    \item What legal barriers currently prevent corporate espionage of AI secrets? How strong are these?
    \item \NC{}: What interventions would help secure algorithmic secrets?
    \item What are the costs of various measures to protect algorithmic secrets? Examples: what research requires \href{https://arxiv.org/abs/2401.14446v1}{what access}~\citep{casper_black-box_2024}, how do SCIF requirements slow down government projects?
    \item If algorithmic secrets have been released, what can be done to limit their spread?
    \item \NC{}, \GOV{}, \DEV{}: AI developers should actually implement good security for algorithmic secrets (possibly aided by government)
    \item[\linkbullet] See \hyperlink{sec:algorithmic_progress}{Algorithmic progress}.
\end{itemize}

\subsubsection[Other than compute and security, what levers exist to control AI development and deployment?]{\impq{} Other than compute and security, what levers exist to control AI development and deployment?} \label{sec:other_levers}

Compute governance and security are typically identified as the main levers for controlling AI development. However, other potentially impactful and currently neglected levers also exist. For instance, human expertise is a critical node---interventions might steer key researchers away from dangerous AI development toward safer activities. Exploring such alternative levers may be particularly important if compute governance alone proves insufficient to ensure a successful Halt.

\begin{itemize}
    \item \NC{}: What approaches can be used to ensure the compute requirements for dangerous AI development remain high?
        \begin{itemize}
            \item What can be done to disincentivize research on algorithmic progress or distributed training? Examples: rules to prevent it, funding, changing social norms.
            \item How can the existing AIs be made less effective at assisting with dangerous AI development? Examples: unlearning and refusal training.
            \item[\linkbullet] See \hyperref[sec:security_algorithmic_secrets]{Security for algorithmic secrets}.
        \end{itemize}
    \item More broadly, what non-compute approaches are available to regulate frontier AI development and support a Halt? Very little work has been done in this space. It is possible that \hyperlink{sec:viability_compute_governance}{compute alone will not be a sufficient governance node}; if that’s the case, what else is possible?
        \begin{itemize}
            \item[\linkbullet] See \hyperlink{sec:visibility_other_projects}{Visibility into other AI projects}.
        \end{itemize}
    \item Are relevant human experts a key node for governance? Alternatively, it is possible humans have been largely obsoleted at some point in the future.
        \begin{itemize}
            \item Does it make sense to modify immigration laws to incentivize relevant experts to come to and remain in coalition countries?
                \begin{itemize}
                    \item[\refbullet{}] \cite{acharya2022comparing, marcopolo_global_nodate}
                \end{itemize}
            \item What can be done to ensure human experts are not working on dangerous frontier AI development? Example: privacy preserving monitoring.
            \item What are historical case studies of getting experts and scientists to avoid certain research? Example: former Soviet scientists incentivized not to work on nuclear technology~\citep{doe1997initiatives}.
        \end{itemize}
\end{itemize}

\subsubsection{Institutional questions}

Many institutions will be needed to implement the desired restrictions, monitoring, and enforcement in a short-term pause. There are many important questions and projects in this category, but we only provide a high-level overview as they are not our area of expertise.

\begin{itemize}
    \item What legal tools could be used to prohibit or restrict certain AI R\&D activities, domestically and internationally?
    \item What institutions or agencies are needed to uphold these restrictions? Successful restrictions include numerous responsibilities, including precise rule-making, monitoring and verification, enforcement, and more.
    \item Which responsibilities can be covered by existing institutions? What institutions need to be created to fulfill other responsibilities?
    \item Which responsibilities can be carried out domestically? Which responsibilities require international governance?
\end{itemize}

\subsubsection{Planning for the longer-term moratorium}

The short-term pause provides an opportunity to establish measures needed for a stable long-term moratorium. Once immediate risks are addressed, we can identify and implement the requirements for a sustained Halt. A key question is determining which decisions must be made now or before initiating the short-term pause, and which can be deferred until after. Careful planning may be needed to avoid locking in a suboptimal trajectory.

\begin{itemize}
    \item What can be punted to the long-term moratorium, and what problems must be addressed during the short-term pause?
    \item How does the world avoid being locked into a situation where a long-term moratorium is harder to implement? For instance, because critical knowledge has proliferated, AI chips have proliferated, bad policies and norms prevent a moratorium, or there is substantial public disapproval of a long-term moratorium.
\end{itemize}

\subsection{Long-term moratorium} \label{sec:long_term_moratorium} 

The Halt is not just an immediate shutdown to avoid imminent harm. Instead, its goal is to last as long as we need, be that years, decades, or centuries. A stable, long-term moratorium may be the only way to buy enough time to solve critical safety problems. A domestic short-term pause could take advantage of the fact that most frontier AI development is in the US. For longer time frames, the Halt needs to be globally enforced. Many of these questions may be answered during the short-term pause when there is substantially more political will and many more people working on these problems. Future AI systems themselves could potentially be useful as well, if we have strong reasons to trust them.

The question of what to do during a long-term moratorium can mostly be pushed to the future. To give some sense, it might involve humanity embarking on an international scientific project to slowly and cautiously develop advanced AI (i.e., a CERN for AI). Some AI activities will be allowed during a moratorium (such as applying narrow AI systems to solve challenges in biology and medicine), and this set of activities will expand as our understanding of AI systems improves. This long-term moratorium would also allow societal adaptation to a world with increasingly advanced AI systems, for instance enabling the world to responsibly handle labor disruption and threats from new weapons technologies.

The Halt could be lifted if humanity has justified confidence that advanced AI development will not cause catastrophe. Reaching this level of understanding of AI systems is a major challenge, likely requiring years or decades of focused effort. There are many known technical safety problems~\citep{yudkowsky2022agi} and likely many that are not yet known. The field currently lacks even a theoretical roadmap for how to confidently solve these problems. There is no guarantee that progress on these problems will happen quickly or predictably.

\begin{itemize}
    \item After a Halt has been initiated, which AI capabilities can be safely developed and deployed widely without imposing unacceptable risks?
        \begin{itemize}
            \item Some AI systems will pose minimal risk and could be run during a global moratorium. How do we know which these are? Examples: based on safety cases, narrow applications. There are many relevant research directions, such as \href{https://arxiv.org/abs/2312.06942}{AI control} schemes~\citep{greenblatt_ai_2024}.
            \item If inference is safe, can secure, inference-specific chips be created to allow physically-widespread access to safe AI systems?
            \item Can chips be created to run specific workloads? Examples: FlexHEGs, signature-requiring hardware.
                \begin{itemize}
                    \item[\refbullet{}] \cite{scher_mechanisms_2024, petrie2024interim}
                \end{itemize}
            \item How could the world change to reduce harm from misaligned and capable AIs? Examples: defensive tech, detecting autonomous behavior, reducing access to AI development resources.
                \begin{itemize}
                    \item[\linkbullet] See \hyperref[sec:defensive_acceleration]{Defensive acceleration}.
                    \item If there are many aligned, defensive-oriented AIs, would they be sufficient to prevent harm from offensive AI systems? How does this balance change as various specific technologies are developed?
                \end{itemize}
        \end{itemize}
    \item What properties of a Halt would allow it to be robust to short-term domestic political changes? Example: it needs to be stronger than the JCPOA (Iran Nuclear Deal).
    \item During a global moratorium, what conditions would safely permit an exception where a highly-secure and monitored AGI project could continue? This may look like an international scientific project (e.g., CERN).
        \begin{itemize}
            \item What multilateral controls would make such development properly democratic?
                \begin{itemize}
                    \item[\refbullet{}] \cite{hausenloy_multinational_2023}
                \end{itemize}
            \item What safety and security precautions need to be taken by such a project?
                \begin{itemize}
                    \item[\linkbullet] See \hyperref[sec:making_usnp_go_well]{Making the US National Project go well}.
                    \item[\refbullet{}] \cite{yudkowsky_six_2022}
                \end{itemize}
        \end{itemize}
    \item Besides deep learning AI development, what other pathways could boost humanity’s intelligence, capabilities, and wisdom in order to first minimize existential risk and then realize a prosperous future? For example, improving human intelligence, pursuing whole brain emulation~\citep{sandberg2008whole}, pursuing other AI development paradigms with better safety properties.
    \item Centralized control of compute resources poses a major risk of authoritarian control. As governments implement a long-term moratorium, how do they avoid authoritarianism while doing the necessary centralization? Examples: MAD, multi-party protocols, trying to teach AIs shared values.
    \item \hypertarget{sec:benefit_sharing}{} How can we enable global benefit sharing? Even without further advancing AI capabilities, there will likely be AI enabled benefits; sharing these benefits could form part of an international agreement. Unlike in the US National Project scenario where the US engages in benefit sharing to placate other nations, in this scenario all actors would be present as more equal participants. 
        \begin{itemize}
            \item What would benefit sharing look like in this context?
            \item What recourse is available if agreements are broken by a party with unilateral control of AI development?
            \item[\refbullet{}] \cite{cankaya_hawkish_2024, okeefe_chips_2024, o2020windfall, heim_ai_2024, justen_sharing_2024, dennis2025options}
        \end{itemize}
    \item If algorithmic progress trends continue, increasing amounts of existing hardware in the world will enable dangerous AI activities. What can be done to avoid this? On the other hand, this may be less of a concern because the difficulty of rogue humans doing algorithmic progress could go up over time. It could go up if ideas get harder to find~\citep{bloom_are_2020}, if the frontier for finding ideas is above human-level (i.e., pushed by above-human-expert AIs), if many years of private R\&D extend the gap between public knowledge and state-of-the-art, or if rogue actors have sufficiently little access to key resources.
        \begin{itemize}
            \item Is it possible to design compliant tech stacks that prevent misuse in a privacy-preserving manner?
            \item[\linkbullet] See \hyperlink{sec:algorithmic_progress}{Algorithmic progress}.
            \item[\linkbullet] See \hyperref[sec:other_levers]{Other than compute and security, what levers exist to control AI development and deployment?}
            \item[\refbullet{}] \cite{grunewald_are_2024, ho_algorithmic_2024, scher_mechanisms_2024}
        \end{itemize}
    \item \F{}: How do we push back breakout time (the time it takes to succeed once efforts start in earnest) for dangerous AI development? Example: dismantling the compute supply chain. How long does breakout time need to be?
\end{itemize}

\clearpage
\section{US National Project} \label{sec:us_national_project} 

The US National Project scenario is a prominently discussed path the world might take as AI capabilities advance, particularly as governments wake up to AI’s national security implications~\citep{aschenbrenner_situational_2024, martin_analysis_2024, USCC2024, zelikow2024defense}. In this scenario, the US Government takes increasingly direct control over domestic AI development while working to prevent other nations from advancing their AI capabilities. This control over domestic AI development could involve centralizing all domestic frontier AI development, championing a particular developer, or something else. The US National Project then accelerates its own AI development, aiming to build AI systems far smarter than expert humans. One potential end goal would be to gain a large military advantage over the rest of the world---a decisive strategic advantage---and use this advantage to reduce AI risks from other countries, potentially while imposing US values globally~\citep{amodei_machines_2024}. This project is unlikely to succeed at its stated goals and is very dangerous. It involves rushing to build advanced AI capabilities, giving minimal time for caution and making it harder to notice problems. We discuss this scenario because it is likely the US Government pursues such a strategy, even though we think it is too dangerous.

If the US Government initiates a US National Project, this will likely alarm other nations. This project could be seen (or even self-described) as a Manhattan Project for AI, aimed at developing powerful weapons which are extremely threatening to global stability. This could spur China to enter the arms race or deter US AI development, for instance by \hyperref[sec:threat_of_sabotage]{threatening to sabotage it}.

Two salient historical analogies are the Manhattan Project and the Apollo Program. Both of these projects involved the US Government spending about half a percent of GDP in their peak funding years, largely with the aim of gaining or displaying dominance over other nations. The Manhattan Project, especially, was started partially due to fears that the geopolitical rival Nazi Germany would win the race to build an atomic bomb.

This section explicitly describes a US National Project, but it is also applicable to situations where government involvement is somewhat weaker. In the next few years, we expect that the US Government will wake up to the possibility of human-level AI systems. At first, this could include efforts to improve security at US AI companies, funding to build AI computing infrastructure in the US, and efforts to reduce proliferation of advanced AI systems to foreign adversaries (e.g., via export controls on advanced AI chips)---some of this is already happening. Later, government involvement could look like a full National Project or some softer version. Softer versions of government involvement might differ in whether AI development is aimed primarily at military dominance, what the overall plan for success is, and the leadership of the projects. The research questions in this section are applicable to scenarios with different levels of government control.

Many of the interventions and research initiatives outlined here benefit both an Off Switch and Halt framework and a potential US National Project. For instance, some measures developed to implement an Off Switch would also help the US maintain a strategic lead in AI. These measures include preventing the spread of frontier AI capabilities via export controls, securing frontier AI model weights, and protecting algorithmic secrets. These same nonproliferation efforts strengthen global readiness to impose a Halt if needed. Research into other actors’ AI capabilities is essential to strategic decision making across all scenarios in this report. Additionally, should a US National Project centralize AI development, subsequent efforts to slow or halt domestic AI progress might be simplified, as coordination would involve fewer distinct AI developers. Finally, if the US National Project decides its AI development strategy is too dangerous, it could potentially pivot into a project focused on designing the technology needed to enforce a halt.
\begin{table}[h]
    \centering
    \caption{The US National Project is vulnerable to numerous failures, many of which would cause catastrophic harm, and all of which would prevent the project from succeeding at its goals. We list these roughly chronologically.}
    \label{tab:usnp_failures}
    \begin{tabularx}{\textwidth}{@{} l >{\raggedright\arraybackslash}X @{}}
        \toprule
        \textbf{Failure Mode} & \textbf{Examples} \\
        \midrule 
        1. Insufficient US lead &
            \begin{minipage}[t]{\linewidth} 
            \begin{itemize}[label=$\rightarrow$, nosep, leftmargin=*, topsep=0pt, partopsep=0pt, itemsep=1pt] 
                \item China surpasses the US in AI capabilities.
            \end{itemize}
            \end{minipage} \\
        \midrule 
        2. Proliferation of dangerous capabilities &
            \begin{minipage}[t]{\linewidth}
            \begin{itemize}[label=$\rightarrow$, nosep, leftmargin=*, topsep=0pt, partopsep=0pt, itemsep=1pt]
                \item Terrorists or rogue states use advanced AI to cause catastrophic harm (e.g., widely accessible biological-weapon AIs).
            \end{itemize}
            \end{minipage} \\
        \midrule 
        3. War &
            \begin{minipage}[t]{\linewidth}
            \begin{itemize}[label=$\rightarrow$, nosep, leftmargin=*, topsep=0pt, partopsep=0pt, itemsep=1pt]
                \item Russia threatens nuclear war if the US attempts an intelligence explosion.
                \item US and China escalate to war because of the threat of an intelligence explosion.
                \item US and China sabotage (or threaten to sabotage) each other's AI progress.
            \end{itemize}
            \end{minipage} \\
        \midrule 
        4. Hardware/software limits &
            \begin{minipage}[t]{\linewidth}
            \begin{itemize}[label=$\rightarrow$, nosep, leftmargin=*, topsep=0pt, partopsep=0pt, itemsep=1pt]
                \item Advancing AI capabilities requires more compute than the US owns.
                \item Returns to automated AI R\&D are not large enough to sustain continual progress.
                \item The current deep learning paradigm fails to scale to human-level intelligence.
            \end{itemize}
            \end{minipage} \\
        \midrule 
        \textbf{5. Misalignment} &
            \begin{minipage}[t]{\linewidth}
            \begin{itemize}[label=$\rightarrow$, nosep, leftmargin=*, topsep=0pt, partopsep=0pt, itemsep=1pt]
                \item \textbf{AI alignment is hard and humanity fails to align advanced AIs; humans lose control.}
                \item \textbf{AI alignment is easy, but development is rushed and humanity fails anyway.}
            \end{itemize}
            \end{minipage} \\
        \midrule 
        6. Failure to develop a decisive strategic advantage &
             \begin{minipage}[t]{\linewidth}
             \begin{itemize}[label=$\rightarrow$, nosep, leftmargin=*, topsep=0pt, partopsep=0pt, itemsep=1pt]
                \item ASI does not break nuclear deterrence (e.g., unable to prevent second strike).
                \item ASI-driven decisive technology development is prevented by adversaries (i.e., sabotage, war).
                \item Actions needed to secure the world are not technologically possible.
             \end{itemize}
             \end{minipage} \\
        \midrule 
        7. Governance failure &
            \begin{minipage}[t]{\linewidth}
            \begin{itemize}[label=$\rightarrow$, nosep, leftmargin=*, topsep=0pt, partopsep=0pt, itemsep=1pt]
                \item Authoritarian power grab via AI.
                \item Failure to use ASI due to concerns about abuses of power.
                \item Actions needed to secure the world are not politically viable (e.g., too invasive).
            \end{itemize}
            \end{minipage} \\
        \bottomrule
    \end{tabularx}
\end{table}
\clearpage

\subsection{Lead-up to the US National Project} \label{sec:lead_up_usnp}

There are actions which are important in the lead-up to a potential US National Project. The US needs mechanisms to ensure its advantage in AI capabilities, including strong export controls and measures to prevent adversaries from obtaining AI intellectual property---such as model weights and algorithmic secrets. Additionally, accurately assessing the US lead is essential, requiring clear visibility into the state of AI models, chip production capabilities, and overall AI progress in other countries.

\subsubsection{How do you ensure the US lead?} \label{sec:ensure_us_lead}

Despite some signs that US companies are currently well ahead in AI development, it is not a given that this lead will be maintained. Maintaining the US lead will involve securing AI compute, model weights, and critical algorithmic secrets~\citep{abecassis2025ai}. At present, algorithmic secrets and potentially model weights are poorly secured, and can readily cross national borders. Additionally, open source models and training techniques will clearly lead to proliferation and help other nations catch up. Below are a series of sub-questions that aim to figure out what it would take to ensure the US stays ahead in the AI race. Note that we do not endorse this strategy of racing to gain unilateral advantage, and we consider it to be highly dangerous.

\paragraph{Export controls} \label{sec:export_controls}
\begin{itemize}
    \item What do successful export controls need to accomplish? What should be the aims of US export controls?
        \begin{itemize}
            \item[\refbullet{}] \cite{heim_understanding_2025}
        \end{itemize}
    \item \NC{}, \GOV{}: How should US Government agencies implement and enforce export controls? Actors within the US Government are best positioned for this question, because they could have influence over export control policy and may have access to privileged information.
        \begin{itemize}
            \item Where are current controls failing at their stated goals? E.g., smuggling pathways.
                \begin{itemize}
                    \item[\refbullet{}] \cite{grunewald2023aichipsmuggling, fist2023preventing}
                \end{itemize}
            \item Where are current export controls going wrong in terms of externalities or unintended consequences? Examples: Maybe algorithmic progress moves faster due to “constraints breeding creativity”, pushing away allies, pushing companies to Chinese chips, incentivizing Chinese indigenization of the chip supply chain (can’t push future on-chip mechanisms), antagonizing China, hurting domestic chip industry.
        \end{itemize}
    \item Other than AI chips and chip manufacturing equipment, what critical AI inputs should be covered by export controls? For example, model weights and algorithmic secrets.
\end{itemize}

\paragraph{What are the trends in compute requirements for frontier AI systems?}
\begin{itemize}
    \item[\linkbullet] See \hyperlink{sec:what_compute_monitored}{What are the trends in compute requirements for frontier AI systems?}
\end{itemize}

\paragraph{Security for AI model weights}
\begin{itemize}
    \item[\linkbullet] See \hyperref[sec:security_ai_model_weights]{Security for AI model weights.}
\end{itemize}

\paragraph{Security for algorithmic secrets}
\begin{itemize}
    \item[\linkbullet] See \hyperref[sec:security_algorithmic_secrets]{Security for algorithmic secrets.}
\end{itemize}

\paragraph{Security to protect against sabotage}
\begin{itemize}
    \item How can an AI project make itself secure against covert sabotage by state actors?
    \begin{itemize}
            \item[\refbullet{}] \cite{harris2025americas}
    \end{itemize}
    \item How can an AI project make itself secure against overt sabotage by state actors (e.g., data center in a bunker)?
    \item How can an AI project make itself secure against emerging technology threats, such as targeted biological weapons?
\end{itemize}

\paragraph{Security to protect against misaligned AI systems}
\begin{itemize}
    \item What security measures could be developed to specifically reduce risks from misaligned AI systems (a form of insider threat)? These might be similar to security improvements targeting AI Control scenarios.
    \begin{itemize}
            \item[\refbullet{}] \cite{greenblatt2025overview}
    \end{itemize}
    \item Advanced AI systems will be able to influence the world in many ways, even without weight exfiltration. For instance, they may be interacting with users over an API, taking agentic actions in the world, or being used internally by AI developers to assist with research. Misaligned AI systems could use these channels, and channels we may not be aware of, to negatively affect the world. Generally, how can an AI project mitigate misalignment risks from known or internal deployment of AI systems?
    \begin{itemize}
            \item[\refbullet{}] \cite{shlegeris2024aicatastrophes, stix2025ai}
    \end{itemize}
    \begin{itemize}
    \item How can an AI project protect its human AI researchers against influence and manipulation by misaligned AI systems? 
    \end{itemize}
\end{itemize}

\paragraph{What other near-term interventions for ensuring the US lead would be useful for the US National Project while also being good across other scenarios?}
\begin{itemize}
    \item What is the viability of increasing immigration to consolidate AI talent? Examples: historical case studies of this, AI immigration trends.
        \begin{itemize}
            \item[\refbullet{}] \cite{acharya2022comparing, marcopolo_global_nodate}
        \end{itemize}
    \item Are there light-touch rules for keeping algorithmic secrets at home? Example: international travel reporting for AGI project employees.
    \item Are there particularly egregious gaps in security, export controls, etc.?
    \item Open sourcing of AI technology poses both a risk to the US lead and a broader proliferation risk. When would it be good to restrict the open release of AI models or research, and how could this be done?
    \item What domestic investment would be best from the standpoint of increasing the US lead without downsides? Examples: funding for safety and security research, chip location tracking technology to support export controls. What investments seem to pose substantial downside risk? Examples: construction of chip fabs, research/training that will be transferred to China.
\end{itemize}

\subsubsection[How do you measure the US lead?]{\impq{} How do you measure the US lead?} \label{sec:measure_us_lead}

To implement a strategy based on maintaining a US advantage, it would be essential to accurately assess the current state of AI progress relative to other nations, especially China. This involves understanding the level of Chinese AI capabilities, including the quality of their AI models, their domestic chip production capacity, and their potential to acquire advanced AI models through espionage or open release. Robust visibility into these areas would provide clearer signals about whether the US lead is secure.

The obvious approach to measuring the capabilities of Chinese models is to perform model evaluations. However, model evaluations face numerous difficulties, including difficulties specific to the adversarial relationship between countries. For example, developers or models themselves might purposefully underperform on evaluations (i.e., sandbagging)~\citep{van2024ai}, developers may “password-lock” their models to be less helpful to others~\citep{greenblatt2024stress}, the field does not know how to upper bound model capabilities in general~\citep{barnett_what_2024, mukobi2024reasons}, and AI capabilities may improve faster than new benchmarks can be created. These issues could make it difficult to evaluate the relative strengths of AI models and projects, but more work could improve this situation. 

\begin{itemize}
    \item Are the current differences in compute capacity likely to ensure US lead in AI development, and over what time scale? For example, this could be interrupted by major algorithmic breakthroughs.
        \begin{itemize}
            \item[\linkbullet] See \hyperlink{sec:what_compute_monitored}{What compute needs to be monitored after a Halt is initiated? More generally, what are the trends in compute requirements for frontier AI systems?}
            \item[\refbullet{}] \cite{EpochMachineLearningHardware2024, hai_global_2024, acharya2022comparing}
        \end{itemize}
    \item What is the state of AI chip production in China?
        \begin{itemize}
            \item How good are the best chips produced in China?
                \begin{itemize}
                    \item[\refbullet{}] \cite{feldgoise_pushing_2024}
                \end{itemize}
            \item How many chips are Chinese fabs able to produce? How long would it take to scale production, under various assumptions?
            \item[\refbullet{}] \cite{grunewald_introduction_2023}
        \end{itemize}
    \item How much can China slow down non-Chinese chip manufacturing? For example, by not exporting important semiconductor raw minerals
    \item What is the state of chip smuggling to China?
        \begin{itemize}
            \item[\refbullet{}] \cite{grunewald2023aichipsmuggling, fist2023preventing}
        \end{itemize}
    \item How likely are Chinese AI projects to obtain model weights of frontier AI systems developed by American companies (e.g., via open release, purchasing, or hacking)?
        \begin{itemize}
            \item What are historical base rates or case studies for spies on key technology projects?
            \item What are the historical base rates or case studies on cyber espionage?
            \item What are the case studies or base rates for state and corporate espionage in the most relevant industries, including AI?
            \item \DEV{}: How good is security around model weights in top AI companies? This is a question for AI developers because they know which security measures they are implementing.
        \end{itemize}
    \item How good are the Chinese models?
        \begin{itemize}
            \item Perform clean comparisons of known models on important benchmarks. Include standard capability benchmarks, agentic task benchmarks, and more qualitative benchmarks.
                \begin{itemize}
                    \item[\refbullet{}] \cite{noauthor_helm_nodate, noauthor_llm_nodate, noauthor_vellum_nodate, chatbot2024leaderboard, EpochLLMBenchmarkingHub2024, scaleseal, wijk_re-bench_2024, details-about-metr-s-preliminary-evaluation-of-deepseek-v3, details-about-metr-s-preliminary-evaluation-of-deepseek-r1}
                \end{itemize}
            \item How good are the algorithmic improvements made by Chinese AI developers? What does this imply going forward?
            \item What problems related to evaluating model capabilities are likely to pose a problem in this context? 
            \item[\refbullet{}] \cite{mukobi2024reasons, barnett_what_2024}
            \item[\linkbullet] See \hyperlink{sec:understanding_forecasting}{Understanding and forecasting model capabilities}
        \end{itemize}
    \item How many resources do Chinese AI companies have?
        \begin{itemize}
            \item How strong are the teams doing AI development at different companies? For example, based on education or previous research.
            \item How much compute do these Chinese AI companies have access to?
            \item How much capital do these companies have?
            \item How reliant are Chinese developers on research and infrastructure from Western developers?
            \item How difficult is it to create or obtain high quality datasets? For example, fine-tuning data or RL environments.
            \item Are AI developers in China or the US substantially hindered by regulation?
        \end{itemize}
    \item \hypertarget{sec:visibility_other_projects}{} Visibility into other AI projects. Many of these would probably be best as US Government projects, for example if they require tracking and surveillance infrastructure, coordination with other nations, or the ability to implement SCIF-level security. 
        \begin{itemize}
            \item \NC{}, \GOV{}: State intelligence gathering about AI development: Build a spy network, signals intelligence (e.g., cyber abilities).
            \item \NC{}, \GOV{}: Track data centers, AI chips, chip production, and other compute infrastructure.
            \item \GOV{}: Is it possible to learn what AI workloads are running in a data center based on various forms of access to that data center? For instance, can you detect training runs by monitoring a nearby power grid?
            \item \NC{}, \GOV{}: Track human experts (e.g., have they disappeared from public or stopped publishing).
            \item OSINT, look at public information to track AI development.
            \item Other verification mechanisms for declared model capabilities.
                \begin{itemize}
                    \item Set up an international whistleblower program.
                        \begin{itemize}
                            \item[\refbullet{}] \cite{noauthor_third_nodate, noauthor_right_nodate}
                        \end{itemize}
                    \item \GOV{}: Set up an international interview program.
                        \begin{itemize}
                            \item[\refbullet{}] \cite{wasil_understanding_2024}
                        \end{itemize}
                    \item \GOV{}: Gain access to models (possibly via espionage) in order to perform evaluations on them.
                    \item \GOV{}: Secure evaluations infrastructure (e.g., create a top security data center dedicated to this).
                        \begin{itemize}
                            \item[\refbullet{}] \cite{team_secure_2024, heim_trusted_2024}
                        \end{itemize}
                    \item[\refbullet{}] \cite{scher_mechanisms_2024}
                \end{itemize}
            \item What other methods could be used to gain visibility into AI projects? Expand this list.
                \begin{itemize}
                    \item[\refbullet{}] \cite{scher_mechanisms_2024}
                \end{itemize}
            \item Broadly, how big is the risk of secret AI projects taking place? What can governments do to detect these? Are there monitoring agreements that would make it more difficult to do a secret project?
        \end{itemize}
    \item How much are countries slowing each other down (e.g., with cyber attacks)?
        \begin{itemize}
            \item Review how much this is happening now (AI and historical).
                \begin{itemize}
                    \item[\refbullet{}] \cite{buchanan_hacker_2020}
                \end{itemize}
            \item What does the AI development attack surface look like and what might be done in the future?
            \item \GOV{}: The US National Project should have an ongoing project to understand the quality and quantity of attacks against it.
            \item What does the offense-defense balance look like for AI security?
                \begin{itemize}
                    \item[\refbullet{}] \cite{garfinkel2019artificial, garfinkel_how_2019}
                \end{itemize}
        \end{itemize}
\end{itemize}

\subsection{Building the US National Project} \label{sec:building_usnp}

There are many ways a US National Project for AI could come to exist. For example, the Government may choose to nationalize domestic AI developers, create its own independent AI project, or implement a public-private partnership. This National Project might look like a Manhattan Project or Apollo Program for AI.

A successful National Project likely involves preventing unauthorized AI development, for instance via absorbing friendly AI projects and blocking adversaries. Key considerations include understanding how a National Project might occur in practice, how centralized control would be managed, and whether the project should remain purely domestic or incorporate international collaboration.

\begin{itemize}
    \item How is the US National Project likely to be created?
        \begin{itemize}
            \item What are the main institutional and legal pathways to creating a US National Project?
            \item What are the most likely methods to create a US National Project for AI? What are the most relevant historical case studies to inform this question? Examples: nationalization, public-private partnerships, government-led coalitions.
                \begin{itemize}
                    \item[\refbullet{}] \cite{cheng2024soft}
                \end{itemize}
            \item Review of historical nationalization and how it might relate to AI.
                \begin{itemize}
                    \item[\refbullet{}] \cite{leung_who_2019}
                \end{itemize}
            \item Forecasting when the government might get involved to various degrees
            \item What would/should leadership for a national AI project look like?
            \item How could such a project be publicly accountable, if this is a goal?
            \item Should we expect this project to have multiple distinct AI projects under its purview, or be a single mega-project? What are the pros and cons of each?
                \begin{itemize}
                    \item[\refbullet{}] \cite{davidson2024shouldtherebejus}
                \end{itemize}
            \item \NC{}, \GOV{}: What does the transition from the current situation to a world with more heavy government involvement look like? Actors within the US Government are probably best positioned to answer this question because they could have more context on the government’s attitude towards intervening in AI development and may have more potential for influence.
            \item If the project starts out as not fully nationalized (e.g., as a public-private partnership), how and why might it transition to being a fully nationalized project? For instance, the project might start out with the aim of maintaining a US lead in AI, and then transition to the full US National Project if the US Government wants to achieve a decisive strategic advantage.
        \end{itemize}
    \item What are the good institutions to carry into this project? Examples: if/then commitments~\citep{karnofsky2024ifthen}, safety teams and research, alignment stress-testing~\citep{hubinger2024introducing}, transparency and accountability~\citep{barnett_bis_2024}, model evaluations~\citep{shevlane_model_2023}. 
        \begin{itemize}
            \item Make these institutions better now so that better versions are carried over.
            \item How could each of these good institutions be carried over?
        \end{itemize}
    \item \hypertarget{sec:bring_in_other_AI_projects}{} \impq{} How could a centralized US National Project bring in other AI development projects (domestic and international)? This is important because centralization could speed up the US National Project and extend its relative lead, enabling more breathing room. Centralization may also be crucial to retaining governmental control of society and the state’s monopoly on violence, due to the effect of advanced AI on soft and hard power. Figuring out how to do such centralization in advance could be especially useful because the situation in the future could be very rushed and chaotic. 
        \begin{itemize}
            \item What are the main institutional and legal pathways a US National project could use to bring in other AI development projects? Example: Title I of the DPA for domestic projects.
            \item Review of the incentives at play in past cases of industry nationalization.
            \item What incentives could affect companies' willingness to join a centralized project (carrots and sticks)? Examples: help with security, pay them, reduce bureaucracy, additional resources (compute, good data), patriotism, legal requirements.
            \item Review past international scientific and military collaborations and learn lessons which could be applicable to AI.
            \item What incentives could affect other countries joining a centralized project (carrots and sticks)? Examples: threats of sanctions, access to resources, access to AI capabilities.
            \item How likely is it for the US National Project to centralize domestic or international frontier AI developers? What are the key factors in such a decision?
        \end{itemize}
    \item How could the US Government prohibit non-compliant projects?
        \begin{itemize}
            \item[\linkbullet] See \hyperref[sec:short_term_pause]{Short-term pause}; many of these interventions can be thought of as levers in an Off Switch.
        \end{itemize}
    \item If this project is international, what other countries are involved and how are they involved?
        \begin{itemize}
            \item What countries are major players in AI development? Where are the major talent pools? Where are the biggest AI data centers?
        \end{itemize}
\end{itemize}

\subsection{Making the US National Project go well} \label{sec:making_usnp_go_well} 

The US National Project may aim to rapidly develop superhuman AI systems via automating AI R\&D (an \textit{intelligence explosion}). This involves running AI systems themselves as automated AI researchers. Carrying out rapid AI capabilities advancement like this is very dangerous because humanity lacks a clear understanding of how to control AI systems and their goals. Therefore, such a project has a significant chance of accidentally developing misaligned AIs that are capable enough to disempower humanity. Developers may not know that they have mistakenly created such a system, as advanced AI systems will be able to fake alignment very convincingly. This is a very dangerous strategy, so we recommend instead pivoting to some form of Halt or Off Switch. If the US Government decides to pursue the US National Project strategy, some interventions could make such a project safer or could make it clear that pivoting is necessary.

Previous \hyperref[sec:ensure_us_lead]{sections} discuss nonproliferation and security efforts that would be crucial to the US National Project’s success---major efforts would be needed to secure it. This section explores questions related to reducing risks through improved institutional design, such as establishing appropriate incentives and leadership structures that encourage caution and prioritize safety. It also covers safety interventions to make running these advanced AI systems incrementally safer. Finally, it addresses methods for minimizing dangerous racing dynamics and avoiding geopolitical conflict.

\begin{itemize}
    \item What institutional features and protocols should this project have to reduce risk?
        \begin{itemize}
            \item What are strategies to ensure project leadership cares about preventing catastrophic risks from AI systems and believes these problems could occur? How do we get explicit acknowledgement of risk from leadership? How do we ensure this project has explicitly benign goals?
                \begin{itemize}
                    \item[\refbullet{}] \cite{yudkowsky_six_2022, christiano_honest_2018,karnofsky_ideal_2022}
                \end{itemize}
            \item \F{}: Does this project have plans and checklists for reducing risk?
                \begin{itemize}
                    \item[\refbullet{}] \cite{bowman_checklist_nodate, karnofsky_playbook_2023,greenblatt2023notes}
                \end{itemize}
            \item What can this project do to ensure commitments are actually followed? Examples: whistleblowers, external verification mechanisms.
                \begin{itemize}
                    \item[\refbullet{}] \cite{christiano_honest_2018, scher_mechanisms_2024}
                \end{itemize}
            \item How does the project have good incentives that allow it to slow down/halt if needed?
                \begin{itemize}
                    \item AGI projects will be under many forms of pressure to move quickly, such as military race pressure and business pressure. Are there contingency plans that can reduce these pressures?
                \end{itemize}
            \item What does a good RSP or If-Then commitment look like, such that following it would entail minimal catastrophic risk? This might look very different from current AI company policies.
                \begin{itemize}
                    \item[\refbullet{}] \cite{metr2025common, karnofsky2024ifthen}
                \end{itemize}
        \end{itemize}
    \item What safety interventions should this project implement to reduce risk?
        \begin{itemize}
            \item \hypertarget{sec:ready_to_go_research}{} \impq{} What ready-to-go research should the US National Project prioritize using AIs for, when AIs are capable of automating AI safety research? This question is important for increasing the likelihood of automated researchers doing useful safety research, and it may be useful to work on this in advance due to the scarcity of expert human labor in such a future. Giving automated researchers well-scoped research projects may increase the likelihood of them doing useful research. 
                \begin{itemize}
                    \item What capabilities are these automated researchers likely to have?
                    \item Which AI safety research directions are relatively automatable? Which can benefit from cheap, low quality labor? Which can benefit from ML engineering?
                    \item What non-AI safety work would be good for the AIs to do? Examples: cybersecurity, forecasting, AI to enable better human cooperation, AI to improve epistemics.
                    \item Collect detailed lists of open problems and research agendas from leading AI alignment researchers. These would ideally include detailed descriptions of which experiments to run.
                        \begin{itemize}
                            \item[\refbullet{}] \cite{greenblatt_comment_2025, nanda_200_2022, goldhaber_building_2025}
                        \end{itemize}
                \end{itemize}
            \item \GOV{}, \F{}: Testing the protocols that are implemented (i.e., fire-drills, whistleblower simulations, wargaming). This is likely a task for people inside the US National Project.
            \item What AI development techniques and methods seem particularly dangerous and should be avoided? For example, allowing AIs to think in neuralese~\citep{andreas2017translating}.
            \item \F{}: What are the threat models to be worried about? What capabilities will these AIs have? How is the project deploying the AIs? What are the resulting threats?
                \begin{itemize}
                    \item[\refbullet{}] \cite{karnofsky2024sketch, shevlane_model_2023}
                \end{itemize}
            \item Pre-training compute will potentially be used as a proxy for danger from AI systems. It is likely insufficient due to algorithmic progress. What can be done to better define existing measures or create better measures? How should FLOPs be counted? How should “effective compute” be calculated? How can data quality be factored in?
                \begin{itemize}
                    \item[\linkbullet] See \hyperlink{sec:algorithmic_progress}{Algorithmic progress}.
                    \item[\refbullet{}] \cite{heim_training_2024}
                \end{itemize}
            \item What role should model evaluations play in our understanding of AI systems and AI risk? What are the key assumptions, drawbacks, and benefits of evaluations?
                \begin{itemize}
                    \item[\refbullet{}] \cite{barnett_what_2024, barnett_declare_2024, mukobi2024reasons}
                \end{itemize}
            \item \hypertarget{sec:safety_plan}{} \impq{} What is a safety plan that would allow an AI project to either successfully build aligned advanced AI, or safely notice that its development strategy is too dangerous? Examples: Safety cases using incapability, then AI Control, then propensity/alignment, then deference~\citep{clymer_safety_2024}. This question is important because better safety plans could avert some catastrophes and lead to a higher likelihood of survival; however, humanity’s understanding of AI systems is sufficiently poor that it may be impossible to construct and follow a sufficiently safe plan in the near future. 
                \begin{itemize}
                    \item What is a safety plan that allows the project to catch bad behavior (rather than building an aligned ASI)? For instance, while automated AI researchers are doing vast amounts of intellectual labor.
                        \begin{itemize}
                            \item[\refbullet{}] \cite{greenblatt_ai_2024}
                        \end{itemize}
                    \item Improve the state of thinking on safety cases early. Prototype safety cases now in order to understand their flaws and improve them, including potentially concluding that a safety case is not feasible in certain situations.
                        \begin{itemize}
                            \item[\refbullet{}] \cite{clymer_safety_2024, balesni_towards_2024, grosse_three_2024, korbak_sketch_2025, goemans_safety_2024}
                        \end{itemize}
                \end{itemize}
        \end{itemize}
    \item How much is the US National Project racing? This is a key consideration for whether they are able to notice danger and successfully solve safety problems.
        \begin{itemize}
            \item What is the state of Chinese AI capabilities?
                \begin{itemize}
                    \item[\linkbullet] See \hyperref[sec:measure_us_lead]{How do you measure the US lead?}
                \end{itemize}
            \item What else might make this project be less race-oriented? Examples: bureaucracy, civilian advisors, public input, beliefs about risks, cautious leadership.
            \item \hypertarget{sec:reduce_racing}{} \impq{} What mechanisms are available to reduce racing between nations? 
                \begin{itemize}
                    \item \NC{}: What verification mechanisms are available to collaboratively reduce racing?
                        \begin{itemize}
                            \item[\linkbullet] See \hyperlink{sec:how_monitor_compute}{How can governments monitor compute they know about, especially to ensure it isn’t being used to violate a Halt?}
                            \item[\refbullet{}] \cite{scher_mechanisms_2024}
                        \end{itemize}
                    \item What trust-based mechanisms are available to collaboratively reduce racing between nations? Examples: Scientific collaboration on safety, demonstrations of misalignment, sharing safety and security research, benefit sharing soft-commitments (as hard commitments are likely technically infeasible). This question is important because actions here could reduce the chance of large-scale conflict or dangerous racing.
                        \begin{itemize}
                            \item Review historical precedent for confidence building mechanisms and see how these might apply to advanced AI development.
                            \item What are the costs of different soft-commitments that an AI developer could make? Examples: dedicated compute, dedicated API tokens, research collaborations.
                            \item Can AI models be developed in a way that signals non-aggression or friendliness? Examples: inclusion of non-aggression in model behavior specifications~\citep{scher_mechanisms_2024, openai_introducingmodelspec_2024} model evaluations for benignness. 
                            \item[\linkbullet] See \hyperlink{sec:benefit_sharing}{How can we enable global benefit sharing?}.
                            \item[\refbullet{}] \cite{brundage_my_2024, karnofsky_playbook_2023, ding_keep_2024}
                        \end{itemize}
                    \item What unilateral mechanisms could reduce racing?
                        \begin{itemize}
                            \item[\linkbullet] See \hyperref[sec:improving_stability_sabotage]{Improving the stability of Threat of Sabotage}. 
                        \end{itemize}
                \end{itemize}
        \end{itemize}
    \item How do the US and China avoid war during this period? A high-level answer might include: benefit sharing commitments, AI defensive technology, MAD dead man’s switch, sharing information to build trust and visibility, or perhaps there is not a major risk of war (maybe war is sufficiently taboo, maybe China doesn’t have sufficient visibility to know the US is building AGI due to US secrecy).
        \begin{itemize}
            \item \F{}: What is the likelihood of conflict in this situation?
            \item How could various actors (e.g., researchers, US Government) make a war less bad? Examples: put compute infrastructure outside population centers, international agreements about new weapons (novel WMDs, inhumane weapons), research clear points where a war could end, defensive tech.
        \end{itemize}
\end{itemize}

\subsection{How does the project end?} \label{sec:usnp_end}

If a nation is following a strategy of racing to superintelligence, it likely envisions some end state, perhaps one where it obtains sufficient military advantage---a decisive strategic advantage---to effectively take over the world. As a reminder, we do not believe this US National Project and decisive strategic advantage strategy are a good approach, but some key decision makers take it seriously~\citep{aschenbrenner_situational_2024, amodei_machines_2024, allen_putin_2017}. This approach involves developing very intelligent AI systems and advanced weapons, both of which could lead to human extinction. Specifically, these advanced AI systems may be misaligned and choose to seize power for themselves. Advanced AI systems will be different from traditional technology in that they will likely have their own goals, so leveraging advanced AI to create a decisive strategic advantage may be a fundamentally flawed strategy.

If the US does undertake this strategy, one ultimate goal should be to stabilize the global order and enable careful reflection on humanity's future. This would require deliberately avoiding authoritarian lock-in and preventing catastrophic outcomes.

\begin{itemize}
    \item \hypertarget{sec:what_is_the_dsa}{} \impq{} What does this decisive strategic advantage look like? This question is important because it may be much safer for an AI project to aim at a specific strategic capability, rather than aiming to build superintelligence and then build some unspecified advanced capability. Aiming at a specific capability could result in keeping AI capabilities lower and incurring less misalignment risk. This question is also crucial to the design and operation of a US National Project, separate from its impact on misalignment risk. 
        \begin{itemize}
            \item What is the overall relatively-safe strategy here? How can we keep the intelligence of the AIs in this scenario low?
            \item What intelligence level is needed for this strategic advantage? How quickly can the project do this? Is there viable retaliation from other countries?
            \item \F{}: What are the decisive weapons and capabilities in this conflict? For instance, is it sufficient to destroy nuclear and AI arsenals?
            \item What are the non-violent ways to attain and execute on this strategic advantage? Examples: Demonstration of capabilities; credible guarantee to benefit sharing; credible guarantee to non-aggression; convince other actors that misalignment risk is high; sufficiently strong defensive technology.
                \begin{itemize}
                    \item Can the project achieve this strategic advantage using a combination of disinformation, propaganda, human manipulation and persuasion? I.e., decisive strategic advantage via super-persuasion rather than physical/cyber attacks or economic advantage.
                \end{itemize}
            \item \GOV{}, \F{}: What is the viability of different methods for establishing strategic control, given the level of AI capabilities? Are cyber attacks sufficient? Questions of military strategy are likely best left to the government.
            \item What does the history of international strong-arming look like, and what might it look like in the AI case?
            \item By default, the project will likely involve automating AI development and then automating other technology developments that enable a decisive strategic advantage. However, AI development itself might be very dangerous (e.g., due to misalignment risk). What safer applications of AI (besides automating AI R\&D) could the US National Project prioritize that would provide strategic advantages while minimizing extinction risks? For instance, AIs that are drop-in replacements for human workers could enable their developer to gain a large economic advantage over other actors without pushing AI capabilities higher, and this could potentially be converted into stable control.
        \end{itemize}
    \item What happens next?
        \begin{itemize}
            \item \NC{}, \F{}: How does the global order change? What does the end state look like?
            \item \NC{}, \F{}: How does the world achieve a state where there is no significant year-on-year risk of human extinction?
            \item \NC{}, \F{}: How can we avoid lock-in or authoritarianism?
            \begin{itemize}
                \item[\refbullet{}] \cite{davidson2025aienabledcoupsho}
            \end{itemize}
            \item \NC{}, \F{}: Figure out what is morally good for the universe. That is, do a \textit{long reflection}~\citep{greaves2020research}.
            \item \F{}: What is the minimal amount of intervention (in other countries) that keeps the world stable? For example, AI enabled surveillance. Given a successful US National Project, this question can be punted to future superintelligent AI systems. 
            \begin{itemize}
                \item[\linkbullet] See \hyperref[sec:long_term_moratorium]{Long-term moratorium.}
                \item[\refbullet{}] \cite{bostrom_vulnerable_2019}
            \end{itemize}
        \end{itemize}
\end{itemize}

We believe that racing through an intelligence explosion is likely to go very wrong: there are many risks, notably the risk of developing misaligned superintelligent AI systems that are uncontrollable. Averting these risks is likely possible in principle, but it could take a substantial amount of research. If key safety challenges need years or decades to solve, racing would make it very difficult to halt for long enough. It is difficult to foresee all of these challenges in advance or estimate their difficulty. Since both noticing and solving problems requires moving slowly, racing sets us up to solve only the easiest problems.

\subsection{Pivoting away from the National Project} \label{sec:pivoting_usnp}

Rather than staying on this dangerous path, it may be advisable to pivot the US National Project substantially and focus on a different goal: a long-term moratorium on dangerous AI activities (development and deployment). This is one of the key points where our thinking differs from others’, such as \href{https://situational-awareness.ai/}{Leopold Aschenbrenner’s Situational Awareness}~\citep{aschenbrenner_situational_2024}. While some advocate for racing through an intelligence explosion, this path appears extremely dangerous and infeasible to execute safely. A less existentially dangerous path forward is a global Halt on dangerous AI activities.

\begin{itemize}
    \item[\linkbullet] See \hyperref[sec:halt_off_switch]{Off Switch and Halt}, especially the section on \hyperref[sec:short_term_pause]{Short-term pause}.
    \item \hypertarget{sec:usnp_realize_too_dangerous}{} \impq{} How does the US National Project realize that its strategy is too dangerous? We believe the US National Project is an unsafe strategy, and insofar as we are correct, it would be useful for the project to learn this fact and pivot to a safer strategy. This question is important because detecting danger may be crucial for averting catastrophe and pivoting to a safer path. 
        \begin{itemize}
            \item[\linkbullet] See \hyperlink{sec:safety_plan}{What is a safety plan that would allow an AI project to either successfully build aligned advanced AI, or safely notice that its development strategy is too dangerous?}
            \item How does the US National Project assess the level of current and upcoming danger? What warning signs or triggers should leadership be looking for?
                \begin{itemize}
                    \item Conceptual work to understand which AI capabilities, propensities, or other traits are most worrisome.
                        \begin{itemize}
                            \item[\refbullet{}] \cite{karnofsky2024sketch, shevlane_model_2023}
                        \end{itemize}
                    \item Empirical evaluations to understand those AI capabilities.
                        \begin{itemize}
                            \item[\refbullet{}] \cite{wijk_re-bench_2024, phuong_evaluating_2024}
                        \end{itemize}
                    \item Forecasting work to predict where those AI capabilities are likely to be in the coming months or years.
                        \begin{itemize}
                            \item[\refbullet{}] \cite{ruan_observational_2024, bhagia_establishing_2024, kaplan_scaling_2020}
                        \end{itemize}
                    \item Model organisms of misalignment.
                        \begin{itemize}
                            \item[\refbullet{}] \cite{hubinger_sleeper_2024, greenblatt_alignment_2024}
                        \end{itemize}
                    \item Honeypots, behavioral red-teaming, and personality evaluations.
                        \begin{itemize}
                            \item[\refbullet{}] \cite{perez_discovering_2022, moore_are_2024}
                        \end{itemize}
                    \item Research on creating and improving propensity / alignment evaluations.
                        \begin{itemize}
                            \item[\refbullet{}] \cite{apollo_we_2024, hubinger_when_2023}
                        \end{itemize}
                    \item Monitoring AI deployment for malicious behavior (outputs and internals).
                        \begin{itemize}
                            \item[\refbullet{}] \cite{greenblatt_ai_2024, macdiarmid2024sleeperagentprobes}
                        \end{itemize}
                \end{itemize}
            \item How does the project determine if AI alignment is hard?
                \begin{itemize}
                    \item What are “easy-mode” analogs to possibly learn from? For example, if model developers can’t get models to be robust to jailbreaking, this is evidence they will also fail to instill more complicated safety properties.
                        \begin{itemize}
                            \item[\refbullet{}] \cite{wang_jailbreak_2024}
                        \end{itemize}
                \end{itemize}
            \item[\linkbullet] See \hyperlink{sec:warning_shots}{What things might serve as warning shots?}
        \end{itemize}
    \item What is the plan for pivoting the US National Project to an Off Switch project?
        \begin{itemize}
            \item What things carry over from the US National Project strategy to the Off Switch world? What things need to drastically change? For example, maybe AI development was already happening in a small number of projects and only these projects need to halt, maybe there is need for a “verification megaproject” to ensure global compliance.
            \item How could one national AI project convince other AI projects from other nations to also stop?
                \begin{itemize}
                    \item[\linkbullet] See \hyperlink{sec:build_common_understanding}{How do we build common understanding about AI risks and get buy-in from different actors to build the Off Switch?}
                \end{itemize}
            \item Institutionally pivoting
                \begin{itemize}
                    \item What is a good chain of command for detecting and responding to danger?
                    \item How can proper incentives be created to allow this project to transition to an Off Switch? For instance, those in power may have to give it up.
                    \item How should the decision to pivot be made? What authority do different actors have over this decision?
                \end{itemize}
        \end{itemize}
\end{itemize}

\clearpage
\section{Other scenarios} \label{sec:other_scenarios}

There are two other key scenarios we think are important to understanding the future of AI development: Light-Touch government involvement, and international Threat of Sabotage. Either of these could occur during the development of advanced AI, but we do not believe either clearly leads to a victory state for humanity---a state where there is low ongoing existential risk from advanced AI systems. However, it may be important for research to focus on these situations because they seem somewhat likely---similar to the above scenarios for the US National Project and the Halt---and there may be useful work in making these scenarios safer or transitioning from them into a safer world.

\subsection{Light-Touch} \label{sec:light_touch} 

The Light-Touch scenario describes a world where the US Government doesn’t heavily intervene in the daily operations or governance of AI companies, similar to today. The US Government may \textbf{assist with or require certain security measures}---without such measures, it appears likely that state actors would steal frontier model weights, substantially reducing any advantage US companies have. The US Government could also attempt to reduce dangerous racing between AI companies by regulating them in various ways. In this scenario, the US Government may contract with AI companies, especially for applying AIs for military and cyber uses. This may look similar to defense contractors that operate largely autonomously while doing substantial business with the government and having specific security requirements.

A Light-Touch approach has the advantage of not alerting or alarming China in the way that a full military US National Project would. Instead, this may look somewhat “business as usual”, with the companies developing increasingly powerful AI systems, and the US Government not giving overt signals that it intends to disempower China. Because the US doesn’t appear to be racing to AGI, China also doesn’t feel compelled to race or sabotage US AI development. The Light-Touch situation also differs from the US National Project in that CEOs or company boards are in charge of AI development, rather than government officials.

The current AI development landscape can largely be thought of as Light-Touch (but without security interventions or interventions to reduce dangerous racing). \textbf{This current situation is untenable} due to AI likely destabilising the global balance of power, threats from human misuse, and risks from misalignment of advanced AI systems. AI systems will meet and surpass human abilities across many domains, often operating faster and at lower cost than humans. Such systems will have massive potential for use in economic and national security activities, such as assisting with the design of novel weapons. These systems will also pose a risk of large-scale harm, such as terrorists using AI systems or misaligned AI systems operating autonomously for their own ends. AI systems this capable will be dual-use; they may enable the development of new dangerous technologies (e.g., biological weapons), and hence will pose major threats to national security. It is difficult to imagine governments not becoming more involved when AI systems begin affecting national and international security. That said, it is worth discussing this scenario because it is the current world and the world many people implicitly assume will continue.

One benefit of a Light-Touch scenario discussed by others~\citep{davidson2024shouldtherebejus} is power decentralization. Compared to a US National Project, the Light-Touch world has more AI projects and therefore might have a more even distribution of power in the world. While this is possible, there are notable uncertainties. First, the current trajectory of AI development appears likely to lead to massive economic inequality, despite there being multiple AI developers. Second, a Light-Touch scenario with many AI developers does not preclude power grabs or the lock-in of bad values~\citep{davidson2025aienabledcoupsho}. It may make such power grabs less likely, but there are still numerous asymmetric tools available to bad actors who wish to grab power, such as extortion, and these would still be feasible in a Light-Touch scenario. While others have expressed that a Light-Touch scenario will lead to a more even distribution of power than other scenarios, we think the net effect is unclear.

\textbf{To our knowledge, nobody has satisfactorily laid out a plan for how the current trajectory can continue without imposing major extinction risk}. That said, there are some potential end states that seem plausible to us: one actor uses advanced AI to obtain a decisive strategic advantage and aggressively institutes unilateral control over AI development, democratic countries use AI for defensive purposes such that other AI development is not a threat, and the international community forms an agreement to control AI activities and prevent catastrophic harm. Below we describe each of these endings in slightly more detail.

First, this situation could end similarly to the US National Project scenario, via the US using a strong military lead (i.e., a decisive strategic advantage) to force other countries to join some agreements (similar to the story discussed by \href{https://darioamodei.com/machines-of-loving-grace#4-peace-and-governance}{Dario Amodei’s Machines of Loving Grace}~\citep{amodei_machines_2024}). That is, this could be similar to the US National Project strategy but with private companies largely in charge, not the government. This strategy has \hyperref[sec:usnp_failures]{similar risks} to that of the US National Project strategy, for example risk of war and risk of misalignment.

Second, this situation could end via US actors developing powerful AI-enabled defensive technology. Using this defensive technology, US actors could prevent adversaries from causing harm with powerful AIs, without directly disempowering them. This ending appears to be what many people imagine when they think about positive AI futures, and it is described by the \href{https://vitalik.eth.limo/general/2023/11/27/techno_optimism.html}{defensive acceleration} (def/acc) movement~\citep{buterin_my_2023}. For this plan to succeed, the defensive technology would need to be developed quickly and would need to be sufficient to prevent harm, even given rapid advances in weapons technology. While possible, we think these assumptions are dubious, and it seems reckless to rely on them. We expect some emerging technologies, such as biological weapons, to have strong offensive use potential while being very difficult to defend against. These technologies would pose an issue for a defense-only strategy. On the other hand, there are some modifications to the def/acc strategy that could improve its viability. For instance, implementing strong controls on the supply chains for potentially dangerous technology, but these would likely require much more government involvement.

Third, the situation could end with an international agreement backed by verification mechanisms, such as the \href{https://www.lawfaremedia.org/article/chips-for-peace--how-the-u.s.-and-its-allies-can-lead-on-safe-and-beneficial-ai}{Chips for Peace}~\citep{okeefe_chips_2024} proposal. In such a scenario, the US and China agree to more visibility between their AI projects and are confident enough that the other will not defect. This is largely due to each side being able to monitor the other’s AI development, and see that they aren’t building an ASI or attempting to achieve a decisive strategic advantage. An international agreement could deteriorate into MAD (nuclear and non-nuclear) or \hyperref[sec:threat_of_sabotage]{Threat of Sabotage} if one side is caught breaking agreements. In this world, the US and China slowly increase the capabilities of their AI systems, while constantly overseeing the other. Eventually they develop ASI, although it is very unclear how this ends. This ending also requires that AI alignment is tractable, such that the continued AI development isn’t catastrophic. An international agreement like this has similar components to a Halt; for example, nonproliferation of dangerous AI capabilities, careful science to determine what development is permissible, and monitoring of AI development. Chips for Peace and related international governance proposals differ from an Off Switch in that they put less emphasis on the capability to halt AI development, a capability we think is critical for preventing human extinction.

All three of these end-game strategies could be pursued simultaneously, as suggested in \href{https://situational-awareness.ai/the-project/}{Situational Awareness} and \href{https://darioamodei.com/machines-of-loving-grace}{Machines of Loving Grace}. Unfortunately, they would still be vulnerable to key failure points. These strategies require that technical AI alignment is a relatively easy problem such that misalignment is somehow avoided during continued AI development. Making an international agreement that prohibits developing misaligned AI, while broadly enabling AI capabilities progress, is currently impossible due to the scientific community’s poor understanding of AI systems---current methods cannot reliably differentiate between these. These strategies also require that defensive technologies are able to secure the world against rogue actors. We think AI alignment could be very difficult, and defensive technologies will not be able to continuously outpace offensive technologies across all relevant domains. Therefore, the current path seems extremely dangerous and will plausibly result in human extinction. Rather than pursue those end-game strategies, it may be better to pivot to a safer strategy, specifically a Halt. If alignment is difficult and companies are moving cautiously enough to \textit{even notice}, they could choose to pivot to this safer strategy. They could then focus on gaining more evidence of the risks and building consensus around which AI development is unacceptably dangerous. This could help them advocate for the required Off Switch infrastructure, and then initiate a Halt if this appears necessary.

\subsubsection{Government involvement} \label{sec:govt_involvement_lt}

While in this scenario the government is not heavily involved in AI development, there will likely be some involvement. For example, it will likely be in the government’s interests to maintain visibility into AI companies, reduce dangerous competition between AI companies, and ensure strong security for AI systems to prevent theft. There are also key questions around how this government involvement could take place, such as which legal frameworks are likely to be applied. There are also currently numerous ongoing efforts to regulate AI development, and it may be possible to steer these positively.

\begin{itemize}
    \item Visibility
        \begin{itemize}
            \item What kinds of transparency should governments have into private AI development? What kinds of transparency should the general public have into private AI development?
                \begin{itemize}
                    \item[\refbullet{}] \cite{belfield_what_2024, casper_black-box_2024, scher_mechanisms_2024}
                \end{itemize}
            \item How can the government ensure developers do model evaluations well?
                \begin{itemize}
                    \item What government levers are available to prevent AI companies from \textit{sandbagging} on dangerous capability evaluations? This could include using minimal resources/effort to elicit performance on dangerous capabilities, giving weaker model versions for external testing, not running relevant tests, not reporting key results, unlearning relevant capabilities, offering refusal-trained models for external evaluation, and more.
                    \item How can model developers set up audit trails or other strong guarantees that deployed models are the same models that are evaluated?
                        \begin{itemize}
                            \item[\refbullet{}] \cite{scher_mechanisms_2024, brundage_toward_2020, millet_aicert_2024}
                        \end{itemize}
                    \item How can model developers do sensitive evaluations without risking leaking classified data? Examples: guarantees of data deletion, allowing models to be evaluated on US Government hardware in secure facilities.
                \end{itemize}
            \item Will AI companies tell the government if they discover evidence that their AI systems are scary or dangerous? Would this be a good thing? What can be done to increase its likelihood?
        \end{itemize}
    \item Competition
        \begin{itemize}
            \item \hypertarget{sec:interventions_coordinate_domestic}{} \impq{} What interventions are available to coordinate the domestic AI projects to reduce corporate race dynamics, while US Government intervention is light-touch? This question is important because such a race could lead to catastrophe. 
                \begin{itemize}
                    \item Options:
                        \begin{itemize}
                            \item Market-oriented interventions, such as a clearinghouse for serving AI models~\citep{ganti2023clearinghouse}.
                                \begin{itemize}
                                    \item[\refbullet{}] \cite{tomei_ai_2025}
                                \end{itemize}
                            \item Safety standards interventions, such as mandated RSPs, universal standards. This could alternatively involve voluntary collaboration from various AI developers to ensure multilateral implementation.
                                \begin{itemize}
                                    \item[\refbullet{}] \cite{larks_rolling_2025}
                                \end{itemize}
                            \item Companies signal their benignness and compliance to each other.
                                \begin{itemize}
                                    \item[\refbullet{}] \cite{brundage_finding_2024}
                                \end{itemize}
                            \item Government inspections, audits, and independent compliance monitors.
                            \item Supply chain restrictions to reduce racing. E.g., central allocation of compute.
                            \item Restrict how many firms can operate.
                            \item Benefit sharing and windfall allocation between companies to reduce winner-takes-all dynamics.
                            \item Companies talk about safety and misalignment problems publicly and with the US Government, and share key research about threats.
                                \begin{itemize}
                                    \item[\refbullet{}] \cite{obrien_coordinated_2024, phuong_evaluating_2024}
                                \end{itemize}
                        \end{itemize}
                \end{itemize}
            \item Without government involvement, how can companies coordinate to reduce racing or share safety and security research? Are there antitrust concerns?
            \item Review of historical case studies on government intervention to reduce race dynamics and improve safety standards, also look at lessons from command economies.
                \begin{itemize}
                    \item[\refbullet{}] \cite{karnofsky_seeking_2023}
                \end{itemize}
        \end{itemize}
    \item \hypertarget{sec:security_lt}{} Security in a Light-Touch world 
        \begin{itemize}
            \item How could AI developers collaborate with the government to improve security for AI models and secrets?
            \item How does the private sector deal with government security integration in existing fields? E.g., in terms of the government providing security assistance or the private sector handling classified information.
            \item[\linkbullet] See \hyperref[sec:security_ai_model_weights]{Security for AI model weights}.
            \item[\linkbullet] See \hyperref[sec:security_algorithmic_secrets]{Security for algorithmic secrets}.
        \end{itemize}
    \item Institutions
        \begin{itemize}
            \item On the default trajectory, what will be the main approaches to governing AI development? What are the pros and cons of these approaches? For instance, some have suggested that existing liability frameworks are likely to be a key part of AI governance~\citep{ball2025ailability}.
            \item Broadly, how does development of national security technology by the private sector typically work? What are the most important lessons to take away from existing public-private partnerships for such technologies?
            \item[\refbullet{}] \cite{zelikow2024defense}
        \end{itemize}
    \item Current regulations. There are light-touch regulations currently in development. How can we steer them to reduce risk? Examples: EU Codes of Practice, various Requests For Comment, voluntary practices from AI companies.
        \begin{itemize}
            \item[\refbullet{}] \cite{barnett_bis_2024, barnett_usaisi_2024, barnett_nist_2024, barnett_romney_nodate, thiergart_omb_nodate, barnett_ntia_nodate, heim_governing_2024}
        \end{itemize}
\end{itemize}

\subsubsection{Proliferation of AI capabilities} \label{sec:proliferation_lt}

It will be important to limit the proliferation of advanced AI capabilities. Proliferation to rogue or incautious actors could lead to large-scale harm such as artificial pandemics or major cyber attacks. Proliferation to foreign adversaries could also weaken any lead US companies have, potentially hurting US economic competitiveness~\citep{BIS2025, heim_understanding_2025}. Nonproliferation efforts will involve \hyperref[sec:export_controls]{export controls} and \hyperlink{sec:security_lt}{security}, as well as other methods.

\begin{itemize}
    \item What dangerous capabilities will the government want to ensure are not widely available? When might these be developed?
        \begin{itemize}
            \item[\refbullet{}] \cite{karnofsky2024sketch, shevlane_model_2023}
            \item[\linkbullet] See \hyperlink{sec:off_switch_proliferation}{What level of AI proliferation would render an Off Switch infeasible?}
        \end{itemize}
    \item Are there methods for training highly capable AI systems so that they can be openly released without posing risks? For example, if they cannot be fine-tuned or jailbroken to do dangerous tasks~\citep{deng2024sophon, tamirisa2024tamper}.
    \item How can we get benefits of widespread access to AI systems without incurring proliferation risks? For example, via API access and light-touch misuse monitoring.
    \item What legal tools could be applied to reduce the proliferation of dangerous AI capabilities or secrets?
    \item What technical tools could be applied to reduce proliferation risk? For instance, Digital Rights Management, Know Your Customer, and improving digital forensics.
\end{itemize}

\subsubsection{Defensive acceleration} \label{sec:defensive_acceleration}

One hope for success in the Light-Touch world is accelerating defensive technologies. If defensive technologies are effective in all relevant domains (i.e., all weapons of mass destruction can be blocked), it may be possible to mitigate misuse risks and Loss of Control risks by constraining the harm that offensive actors can do. We are pessimistic about this approach because we do not expect defensive technologies to be sufficiently effective in all relevant domains---that is a very strong requirement. However, practically it may be possible to avoid some risks by using a defensive acceleration approach.

Notably, this scenario is one in which advanced AI capabilities eventually proliferate to many actors, including rogue actors such as terrorists. It will be difficult to impose retaliatory costs on some of these actors, such as suicide terrorists. Therefore, deterrence regimes such as mutual assured destruction are not applicable, and it is necessary to actually defend against each offensive technology.

\begin{itemize}
    \item Is the combination of light-touch governance and defensive acceleration a viable strategy for avoiding catastrophic risks? This strategy may be hopeless due to some future technologies being much easier to weaponize than to defend against (i.e., their offense-defense balance significantly favors offense). If this strategy is too dangerous, it may be necessary to transition to other strategies.
        \begin{itemize}
            \item[\refbullet{}] \cite{nielsen_notes_2024, bernardi_societal_2025}
        \end{itemize}
    \item What would it look like in the long-term for defensive acceleration to succeed, for instance, eventually leading to a stable multipolar world? What are the biggest barriers, and can they be overcome?
    \item What are the key domains where defensive technology development is needed? What are the domains where defense is most at a disadvantage?
    \item Within the most important domains for defensive technology, what are the most important interventions? Build them.
    \item How can governments and other actors accelerate the development of defensive technology, outside of directly developing it?
    \item Surveillance technologies may prevent harm in a world with broad access to advanced AI capabilities. How could this be done well?
        \begin{itemize}
            \item What are the conditions under which such surveillance could work?
            \item How can surveillance systems be designed to pose minimal risk of abuse or privacy infringement while still reducing large-scale risks? For example, it is likely possible to design a surveillance system with privacy guarantees via using AIs as monitors and wiping their memories.
            \item What are the main things that need to be monitored to address different threats? For example, monitoring DNA synthesis~\citep{crawford2024securing} could reduce some biological risks while imposing only small costs.
        \end{itemize}
    \item What can be done to prevent power grabs by AI companies or AI-enabled actors?
\end{itemize}

\subsubsection{International relations}

Even in a Light-Touch world, the creation of advanced AI systems will still have important geopolitical implications. Advanced AI could lead to the development of technologies that destabilize the balance of power. The economic impacts of advanced AI could also upset existing power structures if, for instance, AI has winner-take-all dynamics. The world may also see a regulatory \textit{race to the bottom} as countries compete to attract AI investment. While there are many interesting questions related to international relations and AI development, we focus on the subset we think are most relevant to catastrophic risks.

\begin{itemize}
    \item What can the US Government or AI companies do to reduce the risk of war or sabotage in a Light-Touch situation?
    \item How would China react to a Light-Touch strategy? Is sabotage from China likely?
        \begin{itemize}
            \item Is there a precedent for nations sabotaging private companies? Historically, how do export controls differ in their effect on the private sector vs. military?
        \end{itemize}
    \item Is an international regulatory race to the bottom likely? How can it be avoided?
    \item How can countries coordinate to reduce proliferation and associated misuse of advanced AI capabilities?
    \item How can companies share safety and security research internationally?
        \begin{itemize}
            \item What are the costs and benefits of sharing different research?
            \item What are the relevant historical precedents?
                \begin{itemize}
                    \item[\refbullet{}] \cite{ding_keep_2024, INAISI2024}
                \end{itemize}
        \end{itemize}
\end{itemize}

\subsubsection{Transitioning away from Light-Touch} \label{sec:transition_lt}

As discussed earlier, the Light-Touch world does not appear tenable in the long term. It may be useful to predict under what conditions there may be political will to transition into an Off Switch and what technical and institutional mechanisms would enable this transition. Similarly, the US government may get involved in AI development, and it could be useful to determine why this would happen. Given that Light-Touch is similar to the current world, the research questions about the \hyperref[sec:halt_off_switch]{Off Switch and Halt} and \hyperref[sec:us_national_project]{US National Project} are broadly relevant here.

\begin{itemize}
    \item Under what circumstances would the US transition away from a light-touch approach and towards a more nationalized project?
        \begin{itemize}
            \item What dangerous capabilities will the government care about solely controlling and when might these be developed? For instance, if \href{https://cdn.openai.com/openai-preparedness-framework-beta.pdf}{Preparedness Level} Critical CBRN is a key capability of interest~\citep{openai_openai_2023}, we may be able to forecast timelines for when this will be developed.
                \begin{itemize}
                    \item[\refbullet{}] \cite{karnofsky2024sketch, shevlane_model_2023}
                \end{itemize}
            \item How much increased lead time does the US gain by doing this? E.g., how much compute would the US have if it aggregated its domestic compute? How else can the US Government prop up domestic AI and push down international competitors?
            \item Do we want to move toward the US National Project strategy sooner or later?
            \item Do the AI company leadership and employees want more government involvement? How might this impact the extent of government involvement?
            \item How does the speed of private projects compare to government/military projects? Review of historical case studies of nationalization, the rate of progress on government vs. private science projects.
            \item[\refbullet{}] \cite{cheng2024soft}
        \end{itemize}
    \item Under what circumstances would the US transition away from a light-touch approach and towards an Off Switch?
        \begin{itemize}
            \item What do warning shots look like in a Light-Touch world?
            \item How can Off Switch infrastructure help alleviate other risks, such that a broader coalition is interested in building an Off Switch?
            \item How could AI companies (rather than the government) advocate for and build an Off Switch?
            \item[\linkbullet] See \hyperlink{sec:build_common_understanding}{How do we build common understanding about AI risks and get buy-in from different actors to build the Off Switch?}
        \end{itemize}
    \item What light-touch government actions can better position the world to later transition into an Off Switch? For instance, export controls, visibility and transparency reporting requirements.
        \begin{itemize}
            \item[\linkbullet] See \hyperref[sec:govt_involvement_lt]{Government involvement}.
        \end{itemize}
\end{itemize}

\subsection{Threat of Sabotage} \label{sec:threat_of_sabotage} 

The Threat of Sabotage regime is characterized by countries threatening to sabotage each other’s AI development in a way that keeps AI capabilities low and prevents dangerous AI systems from coming into existence or being used. A version of this scenario is described in \textit{\href{https://www.nationalsecurity.ai/}{Superintelligence Strategy}}~\citep{hendrycks_superintelligence_2025}, referred to as \textit{Mutual Assured AI Malfunction (MAIM)}, and it may be a default. If sabotaging AI development is sufficiently easy, states could simply do this to prevent their rivals from obtaining highly capable AI systems. States could be concerned with their rivals gaining a strategic advantage via advanced AI capabilities, or developing misaligned AI systems---both of these threats could motivate sabotage or the threat thereof. This would result in a stalling of AI capabilities progress without any international cooperation: instead, the threat of interference causes all parties to keep their AI development to low levels.

This situation is a deterrence regime because the consequences of breaking norms are hopefully large enough to prevent deviation. Nuclear weapons present a relevant historical example of a deterrence regime, and substantial effort has been spent on ensuring the regime’s stability. For instance, the 1972 Anti-Ballistic Missile Treaty limited the use and development of missile defense systems that might undermine nuclear deterrence.

A key aspect of this scenario is countries threatening to sabotage rival AI projects, or otherwise retaliating against norm-violating AI development. Sabotage could follow a ladder of escalation, including cyberattacks, model poisoning, and in more extreme cases destruction of AI data centers or AI chip supply chains. This scenario could involve ascending the escalation ladder via carrying out various forms of sabotage, but it could also result from merely the threat of sabotage, causing countries to self-police and avoid provocative AI development. There are many historical examples of nations sabotaging rival military technology projects. These include US-Israel cyberattacks on Iran’s nuclear program, Allied sabotage of Nazi Germany’s nuclear weapon development during WWII, and US cyber sabotage of North Korea’s nuclear program.

We have considerable uncertainty about whether Threat of Sabotage is a likely regime, what it would look like in practice, and how stable it would be. This regime faces a key challenge: nations must be able to detect and respond to dangerous AI development faster than any single country's breakout time---the time needed to secretly advance beyond permitted levels of AI capabilities. Stability would require that each side has sufficient visibility and sabotage capability. Fortunately, both visibility and sabotage capability are variable and could be increased in order to ensure stability. For example, countries could agree to allow each other certain monitoring abilities to increase visibility. Historically, the Treaty on Open Skies allowed nations to conduct reconnaissance flights over each other’s territory to gain additional confidence about military activities. Nuclear treaties have also been supported by improved monitoring, including both unilateral National Technical Means and multilateral verification.

Other than increasing visibility, countries could extend breakout times by reducing their AI capabilities and infrastructure. For instance, countries could dismantle AI data centers and chip production facilities, similar to historical disarmament such as the nuclear START treaties, and the 1922 Washington Naval Treaty---a treaty in which countries reduced the size of their fleets, including by destroying existing ships. The combination of improving visibility via monitoring and verification, while extending breakout times via limiting AI infrastructure, could make the situation more stable. Given some of these modifications, this regime is similar to a global Off Switch, where countries closely monitor each other’s AI activities and retain the ability to stop dangerous activities.

Even if geopolitical concerns were successfully addressed, this regime would still leave the world vulnerable to misalignment risks. For example, if the US and China were able to somehow credibly commit to not pursue a decisive strategic advantage against each other, they may still recklessly rush to ASI for economic reasons. While this would likely be safer than an international arms race, nations could still build misaligned advanced AI systems, potentially leading to catastrophe. Even if geopolitical tensions can be managed, we need a plan focused on avoiding unacceptably dangerous AI development, such as global Halt.

\subsubsection{How likely is the Threat of Sabotage regime?} \label{sec:likelihood_sabotage}

The Threat of Sabotage regime relies on successful deterrence. In turn, this has numerous requirements and it is unclear if they will be met in the case of AI~\cite{abecassis2025refining}. Deterrence requires clear red lines of AI activity; these should be monitorable, clearly communicated, and related to an unwanted action. For instance, one particular red line could be defined as training a model which reaches some advanced level of cyber abilities. Deterrence also requires that the deterrer credibly has the ability to carry out a threat---such as corrupting an AI training run---and the will to follow through. Deterrence also requires that the threat is sufficiently large and that the deterred party responds to the costs and benefits as expected. Some of these requirements can be broken down into more granular research questions, for instance about the level of visibility, sabotage capability, and breakout time for various dangerous activities.

\begin{itemize}
    \item What activities might serve as a red line or \textit{de facto} capability prohibition? These need to be monitorable, they need to correspond to an unwanted action, and this needs to be communicated between parties. Good answers to this question would enable other work analyzing the stability of this regime.
        \begin{itemize}
            \item[\refbullet{}] \cite{karnofsky2024sketch, shevlane_model_2023, idais_international_2024}
        \end{itemize}
    \item How high will visibility be?
        \begin{itemize}
            \item What are the compute requirements for various AI capabilities of concern? What compute needs to be monitored? E.g., if large pre-training runs are needed, these could be relatively easy to sabotage. But if a small amount of post-training on existing models is needed, it could be very difficult for nations to have early warning and prevent this progress.
                \begin{itemize}
                    \item[\linkbullet] See \hyperlink{sec:what_compute_monitored}{What compute needs to be monitored after a Halt is initiated?}
                \end{itemize}
            \item How can states monitor each other’s AI activities and detect violations of these implicit or explicit agreements?
                \begin{itemize}
                    \item[\linkbullet] See \hyperlink{sec:how_monitor_compute}{How can governments monitor compute they know about?}
                    \item[\linkbullet] See \hyperref[sec:other_levers]{Other than compute and security, what levers exist to control AI development and deployment?}
                \end{itemize}
        \end{itemize}
    \item How high will sabotage ability be?
        \begin{itemize}
            \item What are the key methods countries might use to sabotage each other’s AGI projects? How effective are these? Would these prompt further escalation? Examples: poisoning training data on the internet, interfering with data center power supply, interfering with chip manufacturing, kinetic attacks on data centers.
        \end{itemize}
    \item Is repeated sabotage viable? Potentially the number of sabotage methods decreases to zero as each side strengthens their defenses and uses existing methods; this would make the regime unstable in the long-run.
    \item What is the breakout time for various key activities?
        \begin{itemize}
            \item[\linkbullet] See \hyperlink{sec:compute_breakout}{Compute breakout time}.
        \end{itemize}
    \item Will the security levels in AGI projects be at the required level to enable the Threat of Sabotage dynamic? Threat of Sabotage largely requires that security in the main AGI projects is strong enough to prevent proliferation to non-state actors, but weak enough to enable countries to see each other’s progress and sabotage each other. 
    \item Governments are slow and may not develop the required sabotage methods quickly enough. Are such methods likely to exist in advance? Will countermeasures be implemented in advance?
    \item \GOV{}: What is the historical precedent for international sabotage, both of military projects and private projects, both in and out of war-time? Much of the information here is likely classified, so independent researchers could work on this, but they may not be effective without the right access.
    \item Would world leaders actually pursue aggressive sabotage or other retaliatory actions? Methods like destroying data centers would be a dramatic escalation from current US-China relations, and even historical examples of sabotage have typically occurred during wartime.
    \item Are the retaliatory actions sufficiently costly so as to serve as a deterrent?
    \item What is the escalation ladder? What actions are permissible in response to various provocations?
    \item Based on the answers to the above question, does this regime appear stable?
\end{itemize}

\subsubsection{Improving the stability of Threat of Sabotage} \label{sec:improving_stability_sabotage}

In our view, the stability of the Threat of Sabotage regime depends on three main factors: nations visibility into each other’s AI progress, nations capability to sabotage each other, and the breakout time for dangerous AI activities. There are interventions which could improve each of these, unilaterally or multilaterally. Another crucial component of the Threat of Sabotage regime is the escalation ladder, which governs how nations respond to perceived infractions. Dynamics around the escalation ladder can be improved to increase stability; for example, by ensuring that there are no large jumps in the severity of response, improving communication about which violations will lead to escalation, and by having clear pathways for de-escalation.

\begin{itemize}
    \item Visibility
        \begin{itemize}
            \item What multilateral verification mechanisms could increase visibility?
            \item What unilateral monitoring mechanisms could increase visibility?
            \item \hypertarget{sec:what_is_right_level_security}{} What is the right level of security for AI projects to have? From a visibility standpoint, it may be useful to have relatively weak security, such that countries can spy on each other. Simultaneously, due to proliferation concerns, security should be good enough to prevent other actors from stealing model weights or key secrets. 
        \end{itemize}
    \item \hypertarget{sec:viability_sabotage}{} Sabotage capability 
        \begin{itemize}
            \item How can AI development be exposed to sabotage by the right actors?
            \item How could countries establish sabotage capabilities in a coordinated fashion? What is the palatable and viable version of “putting bombs under each other’s data centers”?
        \end{itemize}
    \item Breakout times
        \begin{itemize}
            \item How could countries verifiably roll back AI infrastructure? What are ways to leverage compute to extend breakout times? For example, dismantle AI data centers, chips, or chip production.
            \item How could countries verifiably roll back AI capabilities to be weaker? If there are AI’s capable of automating AI R\&D, breakout times may be very short, even given limited compute. Therefore, it may be necessary to roll back AI capabilities. Countries might have to delete these systems and revert to weaker AI systems. Verifying that this has taken place appears difficult.
            \item How will algorithmic secrets affect breakout times, and are there methods of governing these secrets that lengthen breakout times?
        \end{itemize}
    \item Escalation
        \begin{itemize}
            \item How can the escalation ladder be improved? For example, what options can be added to it?
            \item How can countries communicate their escalation ladders to each other, insofar as this is desirable?
            \item How could retaliatory actions be narrowly focused and communicated clearly, in order to avoid escalation? For instance, it may be possible to demonstrate that one’s retaliatory strikes would only target data centers, rather than human population centers.
            \item What could de-escalation look like?
            \item What emergency communication channels could be built or used to avoid escalation?
        \end{itemize}
\end{itemize}

\subsubsection{Agreements} \label{sec:agreements_sabotage}

While a Threat of Sabotage regime may come about without any explicit coordination between countries, it may be desirable to have formal or informal agreements as well. Some agreements we are excited about are those that define AI capabilities of concern, coordinate countries around shared interests such as nonproliferation, and preserve optionality for other monitoring agreements. Early agreements to avoid secrecy may be critical to enabling a Threat of Sabotage regime or an Off Switch; without them, countries may not be able to trust each other not to have secret, undetectable, AI development projects.

\begin{itemize}
    \item Should the red lines and deterrence regime be codified via international agreements? What would this look like?
    \item Nonproliferation
        \begin{itemize}
            \item What capabilities should countries agree would be too dangerous if proliferated?
            \item How can countries mutually implement and enforce nonproliferation measures? For example, coordinated export controls, security requirements, restrictions on open source.
            \item[\linkbullet] See \hyperref[sec:proliferation_lt]{Proliferation of AI capabilities}.
        \end{itemize}
    \item Preemptively avoiding secret or hardened AI development
        \begin{itemize}
            \item Are there international agreements that could be made to increase confidence in AI development not happening secretly or in hardened facilities? For example, an international prohibition on putting data centers on submarines or in bunkers.
            \item Could early monitoring agreements reduce the likelihood of secret AI development? For example, if world powers began tracking the location of all AI chips tomorrow, this could make a secret AI datacenter much less likely.
        \end{itemize}
    \item How might nations transition away from the Threat of Sabotage regime into establishing a coordinated Off Switch?
        \begin{itemize}
            \item What warning shots might make nations transition from the Threat of Sabotage regime into an Off Switch? How are misalignment warning shots different from geopolitical destabilization warning shots?
            \item How might interventions implemented for the Threat of Sabotage regime help with the Off Switch? Examples: visibility between nations laying the groundwork for a verification regime, rolling back AI capabilities.
            \item[\linkbullet] See \hyperlink{sec:benefit_sharing}{How can we enable global benefit sharing?}
        \end{itemize}
\end{itemize}

\clearpage
\section{Understanding the world} \label{sec:understanding} 

This section covers questions that are broadly important for understanding the trajectory of AI capabilities, compute requirements, and related trends. These questions are applicable across many different scenarios. They inform our general understanding of how AI technology may develop, how quickly it might advance, and which bottlenecks or breakthroughs could significantly alter the strategic landscape. We also include some high-level strategy questions in this section.

\begin{itemize}
    \item \hypertarget{sec:understanding_forecasting}\impq{} Understanding and forecasting model capabilities 
        \begin{itemize}
            \item \hypertarget{sec:understanding_forecasting}{} This area is important for numerous reasons such as predicting when particular policy interventions are needed, identifying which policy interventions will be effective, and building consensus around when dangerous capabilities will be developed (a point of disagreement that is upstream of many actions). 
            \item \hypertarget{sec:inference_scaling_regime}{} What are the implications of the inference scaling regime? 
                \begin{itemize}
                    \item[\refbullet{}] \cite{epoch2023tradingoffcomputeintrainingandinference, epoch2024optimallyallocatingcomputebetweeninferenceandtraining, heim_inference_2024}
                \end{itemize}
            \item Create high-quality capability evaluations, such as evaluations for learning new skills, and difficult real-world tasks. Evaluations could include static benchmarks, human uplift in controlled settings, and real-world impact studies.
                \begin{itemize}
                    \item[\refbullet{}] \cite{glazer_frontiermath_2024, wijk_re-bench_2024}
                \end{itemize}
            \item Create scaling trends on important benchmarks in terms of relevant input resources. Examples of input resources: compute, time, other capabilities.
                \begin{itemize}
                    \item Attempt to forecast future capabilities with various approaches, such as spending more compute to predict what will be possible with less compute in the future.
                    \item What are the key precursor capabilities or early warning signs for dangerous capabilities? How do these relate to threats~\citep{shevlane_model_2023, barnett_what_2024}? It is currently unclear whether precursor capabilities will always precede and predict dangerous capabilities; this is an underdeveloped research area.
                    \item[\refbullet{}] \cite{ruan_observational_2024, bhagia_establishing_2024, finnveden_extrapolating_2020, kokotajlo2025ai2027, kwa2025measuring}
                \end{itemize}
            \item Looking forward, what are potential bottlenecks for trends in compute use, or ways the current trends might change?
                \begin{itemize}
                    \item[\refbullet{}] \cite{epoch2024canaiscalingcontinuethrough2030}
                \end{itemize}
            \item What is going on with the “spikiness” of model capabilities? Is this a real phenomenon? In what ways are or are not model capabilities profiles spiky?
            \item How will automated AI R\&D affect the development of AI capabilities? What will AI takeoff look like and how quickly will it happen?
                \begin{itemize}
                    \item[\refbullet{}] \cite{davidson_what_2023, davidson2025threetypesofinte, eth2025willairdautomati,johncrox2025takeoff}
                \end{itemize}
        \end{itemize}
    \item What is the state of AI hardware and the computing stock?
        \begin{itemize}
            \item When there are new developments in these areas, should they affect our understanding of the situation? What are the trends we should be expecting? Are new developments consistent with these trends? How should new developments change our thinking?
            \item Who are likely to be the key manufacturers and suppliers of AI hardware in the coming years? For instance, do NVIDIA or TSMC have any notable competition?
                \begin{itemize}
                    \item[\refbullet{}] \cite{EpochMachineLearningHardware2024}
                \end{itemize}
        \end{itemize}
    \item \hypertarget{sec:trends_cost_AI_inference}{} \impq{} What are the trends in the cost of AI inference? Projects could investigate intelligence per dollar, cost of GPUs, efficiencies that make a given model cheaper over time, etc. This question is important for predicting the speed of AI advancements once AI systems are doing a larger fraction of AI R\&D (i.e., takeoff speeds~\citep{davidson_what_2023}), the economic impacts of AI on different timescales, and the rate of algorithmic progress (research on algorithmic progress often focuses on pretraining, but inference is also a key area of interest). 
        \begin{itemize}
            \item What do these trends imply about automated AI R\&D or other notable future developments?
            \item What are the areas where an adversary or malicious actor doing more inference is dangerous?
            \item What do low inference costs imply about synthetic data and capability proliferation? How could AI development practices change if synthetic data is very cheap?
            \item[\linkbullet] See \hyperlink{sec:inference_scaling_regime}{What are the implications of the inference scaling regime?}
            \item[\refbullet{}] \cite{epoch2025llminferencepricetrends, jaipuria_plummeting_2024}
        \end{itemize}
    \item \hypertarget{sec:viability_compute_governance}{} \impq{} How viable is compute governance? Specifically, can the US Government prevent uncooperative foreign actors from gaining sufficient compute to have dangerous models? Can governments in general institute a global moratorium via controls on specialized AI chips? Many questions throughout this research agenda aim to help answer this overall question. This question is important because a large fraction of AI governance research and regulation are premised on compute being a useful node for governing AI, and priorities would need to shift substantially if this assumption turns out to be false. 
        \begin{itemize}
            \item Is China’s access to compute sufficiently constricted? Will future smuggling allow China to be competitive? Taking current trends and extrapolating, will compute restrictions be sufficient, given various timelines and with various modeling parameters?
            \item What are the biggest threats to compute governance and current compute restrictions? Examples: model weight theft, algorithmic progress, distributed training, open source, novel compute supply chains.
                \begin{itemize}
                    \item How do we measure these factors? How do we know if these factors mean compute governance is not viable?
                \end{itemize}
            \item What do researchers, the US Government, and AI developers do differently if compute governance is not viable? Examples: focus on algorithmic secret control, tracking relevant human experts.
                \begin{itemize}
                    \item[\linkbullet] See \hyperref[sec:other_levers]{Other than compute and security, what levers exist to control AI development and deployment?}
                    \item How can we convey this to the relevant decision makers?
                \end{itemize}
            \item How impactful would it be to halt the production of new compute or shift production to only certain chips that have governance controls (e.g., FlexHEG mechanisms~\citep{petrie2024interim})? Existing chips may be sufficient for various activities of concern, thus making such interventions less useful.
            \item The viability of compute governance is likely to fall on a spectrum. What would medium-viability of compute governance look like and how plausible is it? For example, a plausible scenario is that compute requirements to reach dangerous AI capabilities start high but fall quickly.
            \item[\refbullet{}] \cite{sastry_computing_2024, heim_crucial_2024, heim_inference_2024, pilz_increased_2024}
        \end{itemize}
    \item How are different actors orienting to advanced AI development?
        \begin{itemize}
            \item How much are key decision makers in the US and China focusing on advanced AI? Which aspects are they most focused on? For instance, are they interested in economic benefits, military benefits, geopolitical risks, misalignment risks, misuse risks, intelligence explosion implications, or other aspects?
        \end{itemize}
    \item Are there databases or resources to help us and others generally understand the state of AI development?
        \begin{itemize}
            \item[\refbullet{}] \cite{epochai_data_nodate, feldgoise_growing_2024, artificialnalysis_ai_nodate, noauthor_helm_nodate, liang2022holistic}
        \end{itemize}
    \item What high-level plans and strategies for AI governance seem promising?
        \begin{itemize}
            \item What concrete actions would make progress toward successfully enacting each strategy?
            \item What are the major cruxes between different scenarios and strategies? How can these particular cruxes be resolved?
                \begin{itemize}
                    \item[\linkbullet] See \hyperref[sec:appendix_disagreements]{Appendix: Key points of disagreement}.
                \end{itemize}
            \item[\refbullet{}] \cite{aschenbrenner_situational_2024, miotti_narrow_2024, okeefe_chips_2024, martin_analysis_2024, katzke_ai_2024, hendrycks_superintelligence_2025, macaskill2025preparingforthei}
        \end{itemize}
    \item Lists of open research questions in AI governance and related fields.
        \begin{itemize}
            \item[\refbullet{}] \cite{vonknebel_collection_2024, reuel_open_2024, dafoe2018ai, robertson_ai_2024}
        \end{itemize}
\end{itemize}
\clearpage
\section{Conclusion} \label{sec:conclusion} 

There are many difficulties in navigating the development of smarter-than-human AI. Humanity will need to avoid great power war, catastrophic misuse by humans, lock-in of bad values, and most centrally, losing control to misaligned advanced AI systems. We are not ready for these challenges. If we have to face them soon, humanity will probably lose. That is why the thrust of our research is a global Halt on AI. A global Halt is not currently politically tractable, so we focus on preserving the option to halt via building an Off Switch for AI. Rather than incur substantial risk of human extinction, we should take our time, building a deep scientific understanding of AI systems to enable safe development. This research agenda presents our current outlook on the research questions that make progress toward a positive future, and we hope it is a useful resource for others attempting to reduce extinction risk from advanced AI systems.

\section*{Acknowledgements} \label{sec:acknowledgements} 

We are grateful to the following people for useful discussion or feedback. They do not necessarily endorse the content of this document. \\
Ajeya Cotra, Alex Vermeer, Ariel Gil, Brian Judge, Brodi Kotila, Daniel Guppy, David Krueger, Dewi Erwan, Egg Syntax, Erich Grunewald, Erkki Kulovesi, Evan Murphy, Francis Rhys Ward, Harlan Stewart, James Petrie, Joe Collman, Joe Kwon, Joe Rogero, Joshua Clymer, Lily Stelling, Mauricio Baker, Max Räuker, Michael Vermeer, Naci Cankaya, Nora Ammann, Ozzie Gooen, Sarah HW, Vassil Tashev, William Brewer, Yashvardhan Sharma.\\
Special thanks to Malo Bourgon, Lisa Thiergart, Mitch Howe, and David Abecassis.

\addcontentsline{toc}{section}{References}
\bibliographystyle{plainnat} 
\bibliography{references}    

\clearpage

\appendix

\section{Glossary} \label{sec:glossary}

\begin{description}
    \item[AGI] Artificial general intelligence, AI systems that are about as capable as humans across the vast majority of intellectual tasks. For instance, a drop-in replacement for any human remote worker.
    \item[AI alignment] The practice and related research field of ensuring that AI systems have the intended objectives, preferences, and constraints.
    \item[Algorithmic secrets] Techniques and insights that may significantly increase AI capabilities.
    \item[ASI] Artificial superintelligence, AI that can substantially surpass humanity in all strategically relevant activities (economic, scientific, military, etc.).
    \item[Breakout time] The time it would take a country to develop dangerous AI capabilities or do certain AI activities, likely in violation of agreements or monitoring.
    \item[Compute] Refers both to the specialized computer chips used in frontier AI development (e.g., GPUs, TPUs) and to the quantity of operations performed by these chips in AI workloads (i.e., quantity of floating point operations, or FLOPs).
    \item[Compute governance] Using control over specialized AI chips and their production as a lever to govern AI development and deployment.
    \item[Decisive strategic advantage] A position of strategic superiority that would allow one actor (e.g., the US Government) to achieve world domination, where no other actors can effectively challenge them.
    \item[Def/acc] Defensive acceleration, a social movement aiming to accelerate the development of defensive technologies. Generally, accelerating defensive technologies is sometimes viewed as a strategy to counter risks from misaligned AIs or offensive use of AIs, but we are skeptical of its viability.
    \item[Fabs] Semiconductor fabrication plants, these are used to manufacture chips required for AI development.
    \item[Frontier AI] General purpose AI systems which are state-of-the-art, or close to state-of-the-art in their capabilities.
    \item[The Halt] A coordinated and collectively enforced moratorium on dangerous AI development and deployment, lasting as long as is needed to ensure safety.
    \item[Intelligence explosion] AI systems rapidly accelerate AI development, leading to a feedback loop of increasing AI capabilities. The feedback loop could occur at a number of levels: AI software, hardware design, hardware production, or economic.
        \begin{itemize}
            \item \textbf{Software only intelligence explosion} An intelligence explosion where AI systems only accelerate the software side of AI development. This may happen with the current AI hardware stack.
        \end{itemize}
    \item[Light-touch] An AI governance regime where governments allow private sector AI development to proceed with minimal intervention.
    \item[Loss of Control] A potential scenario where one or more AI systems disempower humanity in pursuit of undesirable goals. This refers to humanity losing control of our ability to positively influence the future, rather than losing control of a specific AI system. Loss of Control could involve human extinction.
    \item[Off Switch] The technical, legal, and institutional infrastructure required to shut down sufficiently dangerous AI projects on the global level (i.e., the infrastructure needed for the Halt), if and when it is deemed necessary.
    \item[Proliferation] The spread of dangerous AI capabilities to a wider range of actors, including those who might misuse them.
    \item[Red line] A prohibition or boundary in AI development which actors agree should not be crossed. Certain dangerous AI capabilities, such as the ability to dramatically improve weapon development, are a commonly discussed red line, but there is not yet international consensus on any particular lines. 
    \item[Risk] The combination of likelihood and severity of a harm occurring.
    \item[Safety case] A structured argument that an AI system or development approach meets certain safety criteria or does not pose unacceptable risks.
    \item[Scaling] The practice and paradigm of using increasing computational resources to train more powerful AI models. This is largely driven by empirical AI scaling laws, which demonstrate how AI systems predictably improve (according to training loss) with more resources.
    \item[Threat of Sabotage] A strategic regime where AI capabilities are kept in check due to great powers (i.e., the US and China) sabotaging or threatening to sabotage each other’s AI development.
    \item[Timelines] Estimates of how long until various major AI milestones will be achieved (e.g., AGI or ASI). This is often referred to as “AI Timelines”.
    \item[Uplift] AI systems assisting a human and making them better at a task. For example, assisting a human without biological wet lab experience at synthesizing a virus.
    \item[US National Project] A potential US Government project with the aim of developing advanced AI, potentially well described as a Manhattan Project for AI. The Project may have the goal of eventually developing ASI to gain a decisive strategic advantage over China.
    \item[Verification mechanisms] Mechanisms to confirm compliance with restrictions on AI development and deployment. Some options include on-chip mechanisms, international inspections, and OSINT like satellite imagery.
    \item[Warning shot] An AI emergency, potentially a catastrophic event, which causes the world to become much more aware of the risks posed by advanced AI systems.
\end{description}

\section{Key points of disagreement} \label{sec:appendix_disagreements}

This document is based on a particular worldview and makes a few underlying assumptions, which readers may disagree with. Here we list what we believe are the main cruxes, based on discussion with various stakeholders. If these turned out to be substantially wrong, this would affect our predictions and plans.

\textbf{Feasibility of the Halt}: Most importantly we make two assumptions that the Halt is feasible and worthwhile compared with other options:
\begin{itemize}
    \item \textbf{Political feasibility}: That it is possible to achieve sufficient international consensus for a coordinated, long-term Halt on dangerous AI development.
    \item \textbf{Technical feasibility}: That effective mechanisms can be implemented to enforce the Halt. These could include: compute governance, monitoring and verification mechanisms, and mechanisms to restrict certain types of AI research.
\end{itemize}

\textbf{Relative importance of the Halt}: We believe non-Halt strategies to develop advanced AI systems impose major risks, and it is difficult to alter these paths such that risk is substantially reduced. Our focus on the Halt is due to a combination of the likelihood of different scenarios happening and how successful we think work in these scenarios would be at reducing catastrophic risks. There is useful work to improve other strategies, but we are overall less excited about this other work because we do not expect it to substantially reduce risk.

\textbf{AGI and ASI are possible and powerful}: It is feasible for humanity to eventually develop AI systems that match and exceed human performance across all intellectual domains. Such systems would have a massive impact on the world, similar in scale to the agricultural or Industrial Revolution.

\textbf{Timelines and takeoff}: We believe AI systems capable of posing catastrophic risks will likely be developed within years to decades, rather than centuries. Additionally, capabilities will increase rapidly, going from around human-level to superhuman in less than a decade (and potentially much faster).

\textbf{Alignment difficulty}: We assume that AI alignment is a hard technical problem, and that misaligned advanced AI systems pose substantial risks of loss of control and potentially human extinction.

\textbf{Defensive technology limitations}: We believe that defensive technologies will often be insufficient to defend against AI-enabled offensive technologies, particularly for threats like biological weapons. This makes us much less optimistic about plans aiming to achieve a stable multipolar world or plans which allow for the widespread proliferation of powerful AI models. This applies for both technology to defend against terrorists and technology which would allow nations to maintain a defensive advantage against other nations. We are concerned about catastrophes even if they fall short of total human extinction.

\textbf{US government strategic awareness}: We assume the US government will eventually take the national security and economic repercussions of advanced AI seriously. AI will eventually be a major focus of governments.

\textbf{Chinese strategic awareness}: We assume China will wake up, and recognize and respond to US AI development, rather than remaining passive while the US develops ASI unhindered.

\textbf{No one wins a race to ASI}: In a race to ASI, China achieving ASI first is similarly bad as the US, because the ASI will likely be misaligned. A misaligned ASI would cause extinction, regardless of who created it.

\textbf{Feasibility of a decisive strategic advantage}: We assume achieving a decisive strategic advantage is neither trivially easy (i.e., not requiring advanced AI capabilities) nor so hard that the US National Project could never work.

\textbf{Likelihood of conflict}: We assume that the development of advanced AI could easily lead to conflict. Specifically:
\begin{itemize}
    \item \textbf{The US might attempt a decisive strategic advantage}: If it were capable, the US might attempt to develop AI systems to achieve a decisive strategic advantage. At a minimum this would include control over international AI development, and it could involve imposing US values globally.
    \item \textbf{China might threaten war in response}: If China was aware that the US was building ASI and attempting to achieve a decisive strategic advantage, then China might threaten war.
\end{itemize}

\textbf{Likelihood of the various scenarios}: Overall, we believe that the scenarios we describe will somewhat resemble what happens. If we thought a particular scenario, or variation of it, was very unlikely to transpire, we would not focus on such a scenario.

\end{document}